\documentclass[a4paper,fleqn,usenatbib]{mnras}

\usepackage{newtxtext,newtxmath}

\usepackage[T1]{fontenc}
\usepackage{ae,aecompl}

\usepackage{graphicx}	
\usepackage{amsmath}	
\usepackage{amssymb}	
\usepackage{multirow}

\title[Photometric study of the SMCNOD]{Photometric study of the SMCNOD using variable stars from the OGLE-IV survey}
\author[Prudil et al.]{
Z. Prudil$^{1}$\thanks{E-mail: prudilz@ari.uni-heidelberg.de}, E. K. Grebel$^{1}$, I. D\'ek\'any$^{1}$, R. Smolec$^{2}$ \\
$^{1}$ Astronomisches Rechen-Institut, Zentrum f{\"u}r Astronomie der Universit{\"a}t Heidelberg, M{\"o}nchhofstr. 12-14, 69120 Heidelberg, Germany\\
$^{2}$ Nicolaus Copernicus Astronomical Center, Polish Academy of Sciences, ul. Bartycka 18, 00-716 Warszawa, Poland\\
}

\date{Accepted XXX. Received YYY; in original form ZZZ}

\pubyear{2017}

\begin{document}

\label{firstpage}
\pagerange{\pageref{firstpage}--\pageref{lastpage}}
\maketitle

\begin{abstract}
We present a study of a recently discovered stellar overdensity near the northern edge of the Small Magellanic Cloud (SMCNOD). We exploited variable stars from the fourth release of the Optical Gravitational Lensing Experiment (OGLE). We used mainly pulsating variable stars and investigated their potential association with the SMCNOD using their spatial distribution and distances. We found four rather spatially concentrated anomalous Cepheids and eight evenly dispersed RR~Lyrae stars to be most likely members of this overdensity. The anomalous Cepheids inside the SMCNOD trace possible intermediate-age population with ages ranging between $2-4.5$\,Gyr. The age distribution of anomalous Cepheids seems to be in a good agreement with the age distribution of anomalous Cepheids in the SMC. Using empirical relations for RR~Lyrae stars we determined the median metallicity for a possible old population in the SMCNOD to be $\rm [Fe/H]_{\rm SMCNOD}=-1.71\pm0.21$\,dex, which is in agreement with median metallicity of the old SMC population. The density profile for anomalous Cepheids shows a small anomaly at the position of the SMCNOD, on the other hand, RR~Lyrae variables show no such deviation. The probability of finding the observed number of variable stars at the location of the SMCNOD by chance is very low for anomalous Cepheids (0.7\,\%) but high for RR~Lyrae stars (13.0\,\%). Based on its variable stars content, we thus confirm the presence of a modest overdensity in intermediate-age stars in the SMCNOD and conclude that it probably has its origin in the SMC rather than to be the remnant of an accreted dwarf galaxy.
\end{abstract}
%



\section{Introduction}

The Magellanic System is an important nearby cosmological laboratory. It consists of the Large and the Small Magellanic Cloud (LMC and SMC), and the Magellanic Bridge and Stream that were most likely formed through the interaction between both Clouds and possibly the Milky Way \citep{Gardiner1994, Gardiner1996, Connors2006, DiazBekki2011, DiazBekki2012, Onghia2016}. In addition, the Magellanic system may include a number of ultra-faint dwarf spheroidal galaxies \citep[e.g.,][]{Bechtol2015, Drlica2015, Drlica2016}. The SMC is a smaller counterpart of the LMC with an estimated mass between $\rm 1 - 5 \times 10^9\,M_{\odot} $ \citep{Kallivayalil2006}. The SMC is an irregular, star-forming, low-metallicity, gas-rich dwarf galaxy. It has an elongated shape. Its northern part is younger and closer to us, while the south-western part is older and more distant \citep{Haschke2012bb, JD2016CCs}. The SMC consists of several substructures, e.g., the Wing \citep{Cioni2000} which together with the central region shows the most active star-formation in the SMC.

Recently, a small stellar overdensity, 8$^{\circ}$ north of the center of the SMC, was found by \cite{Pieres2016SMCNOD}. They used data from the Dark Energy Survey \citep[DES,][]{TDES2005} and the MAGellanic SatelLITEs Survey \citep[MagLiteS,][]{Drlica2016} to trace and study this overdensity. \cite{Pieres2016SMCNOD} called this feature the Small Magellanic Cloud Northern Over-Density (SMCNOD). According to \cite{Pieres2016SMCNOD} it contains mainly intermediate-age stars ($\sim 6$\,Gyr with a metallicity of $\rm Z=0.001$) with a small fraction of young stars ($\sim 1$\,Gyr, $\rm Z=0.01$). The center of the SMCNOD is estimated to be located at $\rm R.A._{J2000}=12.00$\,deg and $\rm Dec_{J2000}=-64.80$\,deg, and the structure has an ellipticity of $\epsilon=0.6$ with a position angle of 350.9$^{\circ}$. Based on its colour-magnitude diagram (CMD), \cite{Pieres2016SMCNOD} estimate the distance to the SMCNOD to be the same as the distance to the SMC \citep[i.e., a distance modulus of $(m - M)_{0}^{\rm SMC} = 18.96 \pm 0.02$,][]{GrijsBono2015}. 

\citeauthor{Pieres2016SMCNOD} also discuss several possible origins of the SMCNOD. One involves a past LMC - SMC encounter, which would result in an enhanced star formation activity in both objects (SMC and LMC) and imply that the SMCNOD consists of material that was tidally stripped from the SMC. The second possibility involves resonant stripping in dwarf-dwarf encounters \citep{Onghia2009} which might lead to the formation of shells comparable with the SMCNOD. The third hypothesis suggests that the SMCNOD is a small dwarf galaxy currently orbiting around and/or being tidally disrupted by the SMC. This would imply a fairly low metallicity and a narrower age distribution for the SMCNOD in comparison with the SMC.

In this paper, we explore the nature of the SMCNOD using variable stars. We present an analysis of photometric data of variable stars from the Optical Gravitational Lensing Experiment IV \citep[OGLE-IV,][]{Udalski2015-OGLE}. In \S~\ref{sec:DataAnalysis} we study the SMCNOD using variable stars of various types located in the range of the coordinates of this apparent feature. Wherever possible we determine stellar distances, masses, ages, and metallicities. We discuss the spatial distribution of the variable stars and their metallicities in comparison with the main body of the SMC. Section~\ref{sec:FieldStaAnalysis} contains a field star analysis where we divided the northern outskirts of the SMC into small ellipses with the same proportions as the SMCNOD. Furthermore, we discuss differences between the variable stars in these selected comparison fields and the variable stars inside the SMCNOD. Whether the variable stars at the location of the SMCNOD represent a significant overdensity is explored in \S~\ref{sec:DLS}. In Section \ref{sec:Conclus} we summarise our findings. The electronic version of the supplementary material contains information about the analyzed stars and their parameters.

\section{Data analysis} \label{sec:DataAnalysis}

For the purpose of this paper, we have explored the latest release of OGLE-IV data of the Magellanic Clouds \citep{SoszynskiLeavit2017}. The OGLE observations are conducted in the $I$ and $V$-passbands of the Johnson-Cousins photometric system. Apart from the processed photometric data, additional information about variability is also made available (e.g., periods of variation, epochs, peak-to-peak amplitudes, and Fourier coefficients). We used publicly available data for variable stars in OGLE-IV for the SMC to study the recently discovered SMCNOD \citep{Pieres2016SMCNOD}. In the following subsections, we briefly discuss the different types of variable stars used in our study.

\subsection{Eclipsing binaries} 

Eclipsing binaries can be invaluable tools. They can serve as laboratories for testing stellar evolution theory and for constraining the rate of mass transfer and mass loss. Binary systems offer a possibility to measure directly stellar properties like mass, temperature, radius, surface gravity, and absolute magnitude. Perhaps most importantly, in the context of stellar population studies, they can serve as distance indicators in the Milky Way and nearby stellar systems \citep{Pietrzynski2013, Graczyk2014SMC}.

\begin{figure} 
\includegraphics[width=\columnwidth]{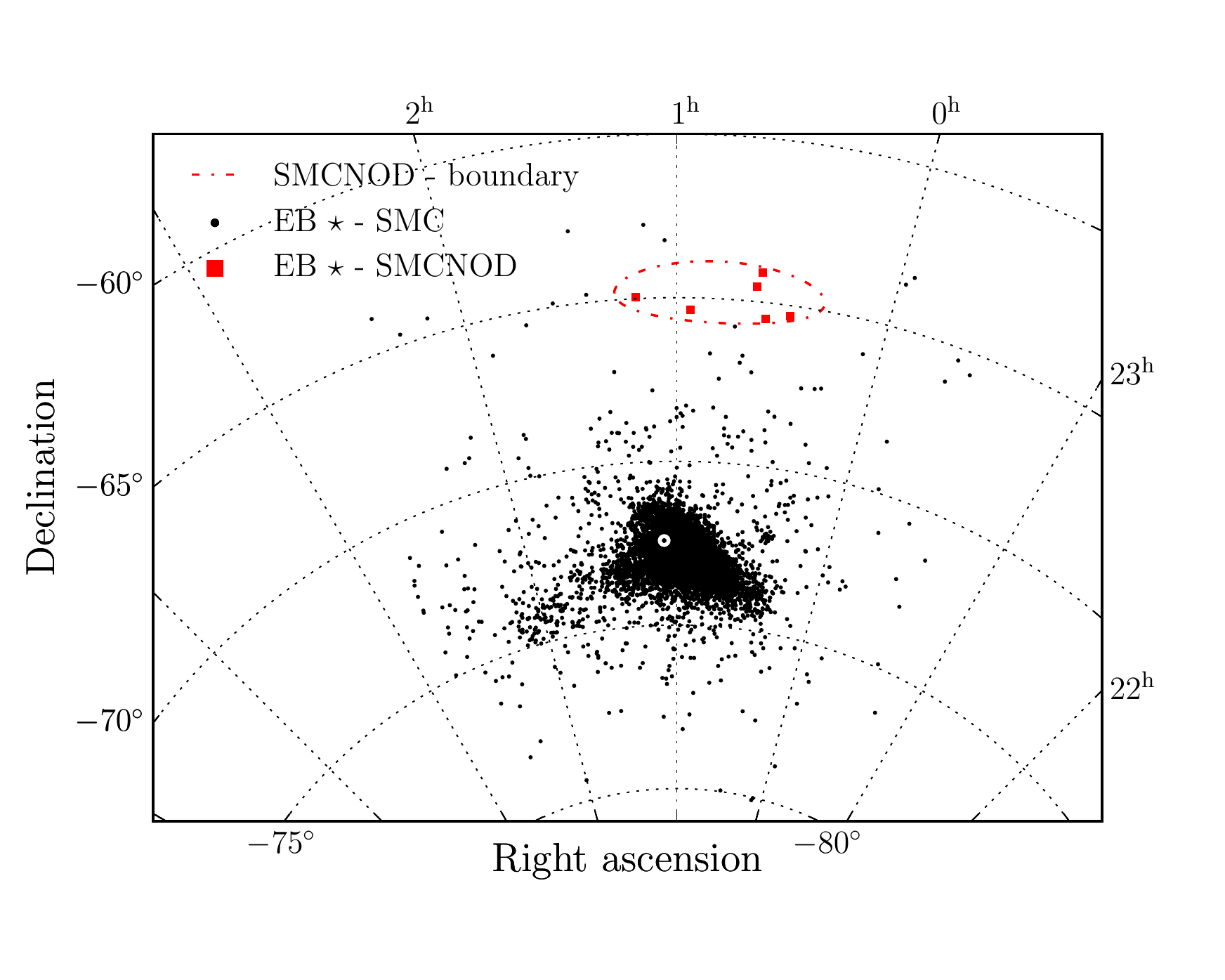}
\includegraphics[width=230pt]{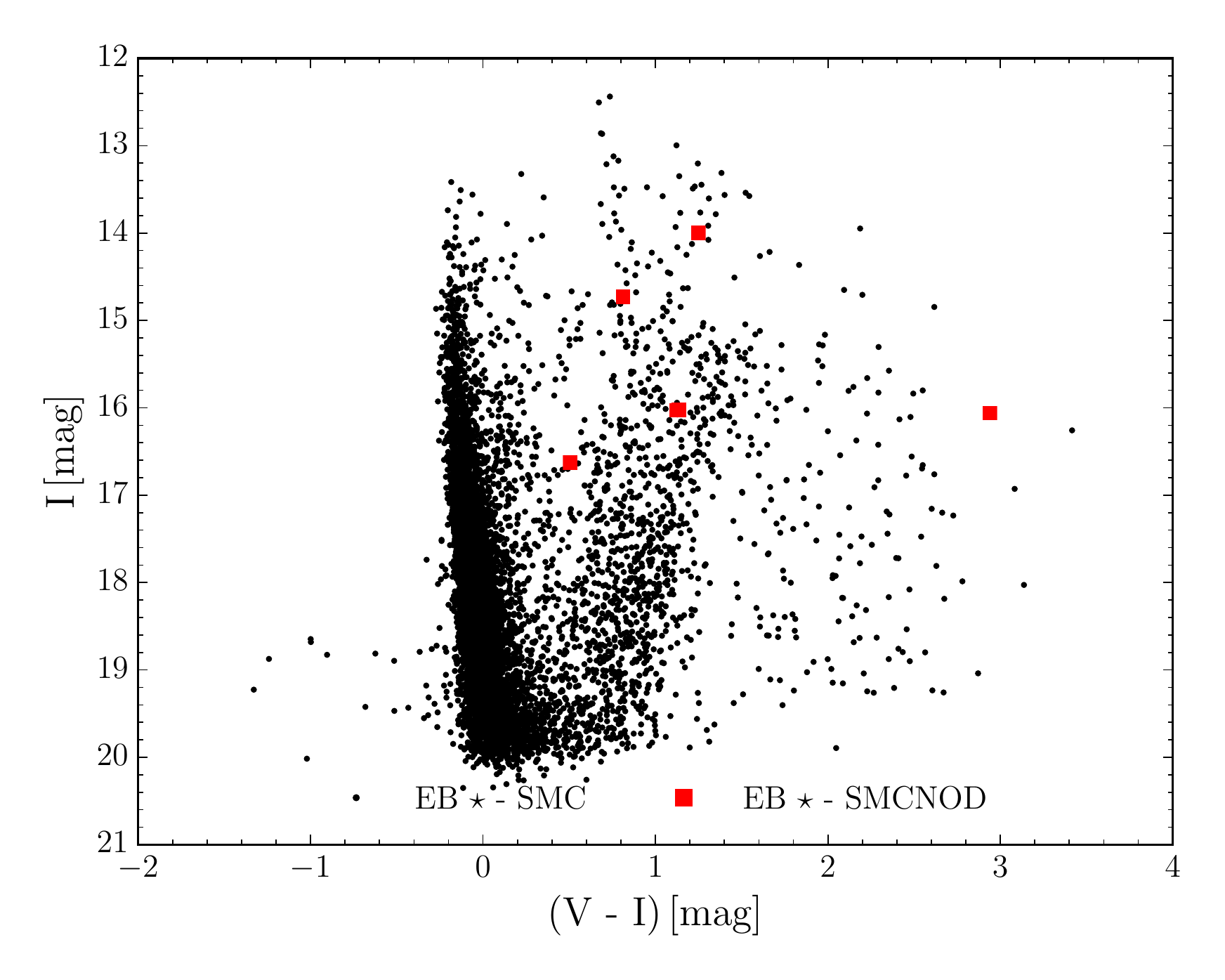}
\caption{Spatial distribution and CMD for eclipsing binaries (EBs) in the SMC. The top panel shows the distribution of EBs in equatorial coordinates for the SMC (black dots). Red squares represent EBs that lie at similar coordinates as the SMCNOD. The red dash-dotted ellipse indicates the approximate position of the SMCNOD. The white circle denotes the apparent kinematic center of the SMC \citep{Stanimirovic2004}. The bottom panel displays a CMD of the EBs seen towards the SMC with the same color coding as in the upper panel.}
\label{fig:SMCNOD-EC}
\end{figure}

The OGLE-IV data release for eclipsing binaries (EBs) contains over 48\,000 objects in the Magellanic system \citep{Pawlak2016ECBIN}, of which 8\,401 belong to the SMC. The top panel of Fig.~\ref{fig:SMCNOD-EC} shows their distribution in equatorial coordinates, highlighting those 6 EBs whose positions overlap with the SMCNOD.

In the bottom panel of Fig.~\ref{fig:SMCNOD-EC} we plotted the CMD of the EBs found in the SMC. The six binaries that overlap with the SMCNOD do not have any $V$-band OGLE-IV photometry. Therefore, we used $G$-band magnitudes from Gaia DR--2 \citep{Gaia2018}, for the aforementioned EBs. Using the coefficients from table 3 in \cite{Jordi2010}, we transformed these $G$-magnitudes into the $V$-band and used those in the CMD in Fig.~\ref{fig:SMCNOD-EC}.

The basic properties of these six binaries can be found in Table~\ref{tab:SMCNOD-EC-table}. We note that they all belong to the eclipsing detached/semi-detached subtype. The binaries found at the position of the SMCNOD fall on the brighter end of the $I$-band magnitude distribution. The shape of their light curves suggests that they may belong to the group of $\beta$\,Persei, late-type eclipsing systems. From these six binary systems, four seem to have similar $I$-band magnitudes, in addition, two of them share almost the same location in the CMD (OGLE-SMC-ECL-6334 and OGLE-SMC-ECL-6349, see lower panel of Fig.~\ref{fig:SMCNOD-EC}). Under the assumption that they lie inside the SMCNOD their approximate mean absolute magnitude would be $\simeq -3$\,mag \citep[assuming that the SMCNOD and the SMC lie at the same distance,][]{Pieres2016SMCNOD}. This would be in agreement with their evolved state inferred from their CMD locus (post-main-sequence). Detached evolved systems with longer orbital periods have been found and used to study the structure of the Milky Way and to determine the distance to the SMC \citep[see,][]{Helminiak2013,Graczyk2014SMC}. To determine individual distances we would need additional information about the studied systems, e.g. near-infrared photometry, radial velocity curves, or light curve modeling. Unfortunately, neither of these is currently available. Thus, the association of the four aforementioned binaries with the SMCNOD remains tentative.


\setlength{\tabcolsep}{5pt} 
\begin{table}
\centering
\caption{Basic properties of the six eclipsing binaries that lie at the same coordinates as the SMCNOD. Column 1 contains the name of the star in the form OGLE-SMC-ECL-ID, but only the ID number itself is listed here for brevity. Columns 2 and 3 contain equatorial coordinates (J2000). Mean $I$-band and $V$-band magnitudes can be found in columns 4 and 5, respectively. Column 6 contains orbital periods.}
\label{tab:SMCNOD-EC-table}
\begin{tabular}{lccccc}
\hline
ID           & R.A.\,[deg] & Dec\,[deg] & $I$\,[mag] & $V$\,[mag] & $P$\,[day] \\ \hline
6268 & 6.93404       & -65.31314     & 16.627 &   17.133  & 1.3383077   \\
6319 & 8.65396       & -65.49322     & 14.731 &  15.543   & 3.6228675   \\
6334 & 9.18079      & -64.10989     & 16.028 &  17.152   & 1.5717865   \\
6349 & 9.44271       & -64.54739     & 16.021 &  17.157   & 4.0081439   \\
7162 & 14.01704      & -65.36956     & 13.998 &  15.247   & 0.8175277   \\
7856 & 17.87579      & -64.95067     & 16.061 &  19.000   & 1.9892185  \\\hline
\end{tabular}
\end{table}

\subsection{Classical Cepheids}

Classical Cepheids (hereafter referred to as CCs) are radially pulsating Population I variables with periods ranging from a few hours up to several weeks. They are well-established distance indicators in the Milky Way and in extragalactic systems.

\begin{figure} 
\includegraphics[width=\columnwidth]{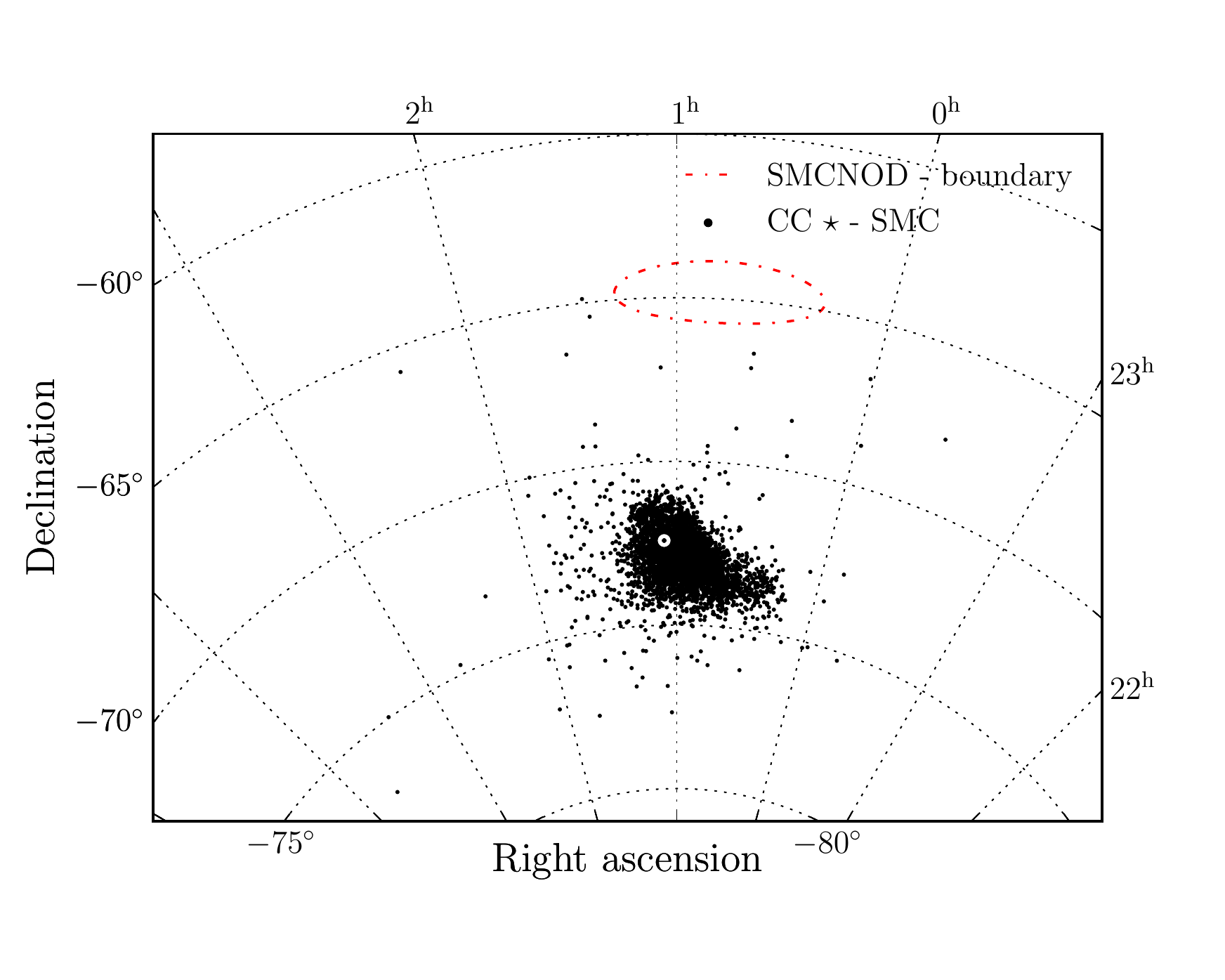}
\caption{Similar as Fig.~\ref{fig:SMCNOD-EC} but for classical Cepheids in the SMC.}
\label{fig:SMCNOD-CC}
\end{figure}

Thanks to their continuing star formation activity the SMC and LMC contain a large number of CCs \citep[e.g.,][]{Haschke2012bb,Haschke2012a}. In the OGLE-IV photometry, we can find 9\,649 CCs, of which almost 5\,000 belong to the SMC \citep{Soszynski2015CCs}. Fig.~\ref{fig:SMCNOD-CC} shows the celestial distribution of CCs in the SMC. None of the CCs overlap with the SMCNOD. Most of them are concentrated in the central regions with very few located in the outskirts of the SMC \citep[see, e.g.,][]{Haschke2012bb}. The lack of CCs in the SMCNOD suggests a lack of stars younger than a few hundred million years.

\subsection{Anomalous Cepheids} \label{sec:AC-dista-klasika}
Anomalous Cepheids (hereafter ACs) are radial pulsators with rather short pulsation periods (ranging from a few hours up to two days). The ACs are believed to be relatively massive \citep[up to 2~$\rm M_{\odot}$,][]{Marconi2004}, metal-deficient helium-burning stars. They are often found in nearby dwarf galaxies \citep[see][]{Kinemuchi2008, Bernard2009}, but their occurrence in globular clusters is very rare with one confirmed case \citep{Zinn1976}. The origin of ACs is still unclear; there are two main hypotheses to explain them. They may either be metal-poor intermediate-age (1 - 6\,Gyr) stars or merged binary stars older than 10\,Gyr. Importantly, they follow a P-L relation and can be used as standard candles \citep{Soszynski2015ACs}.

\begin{figure}
\includegraphics[width=\columnwidth]{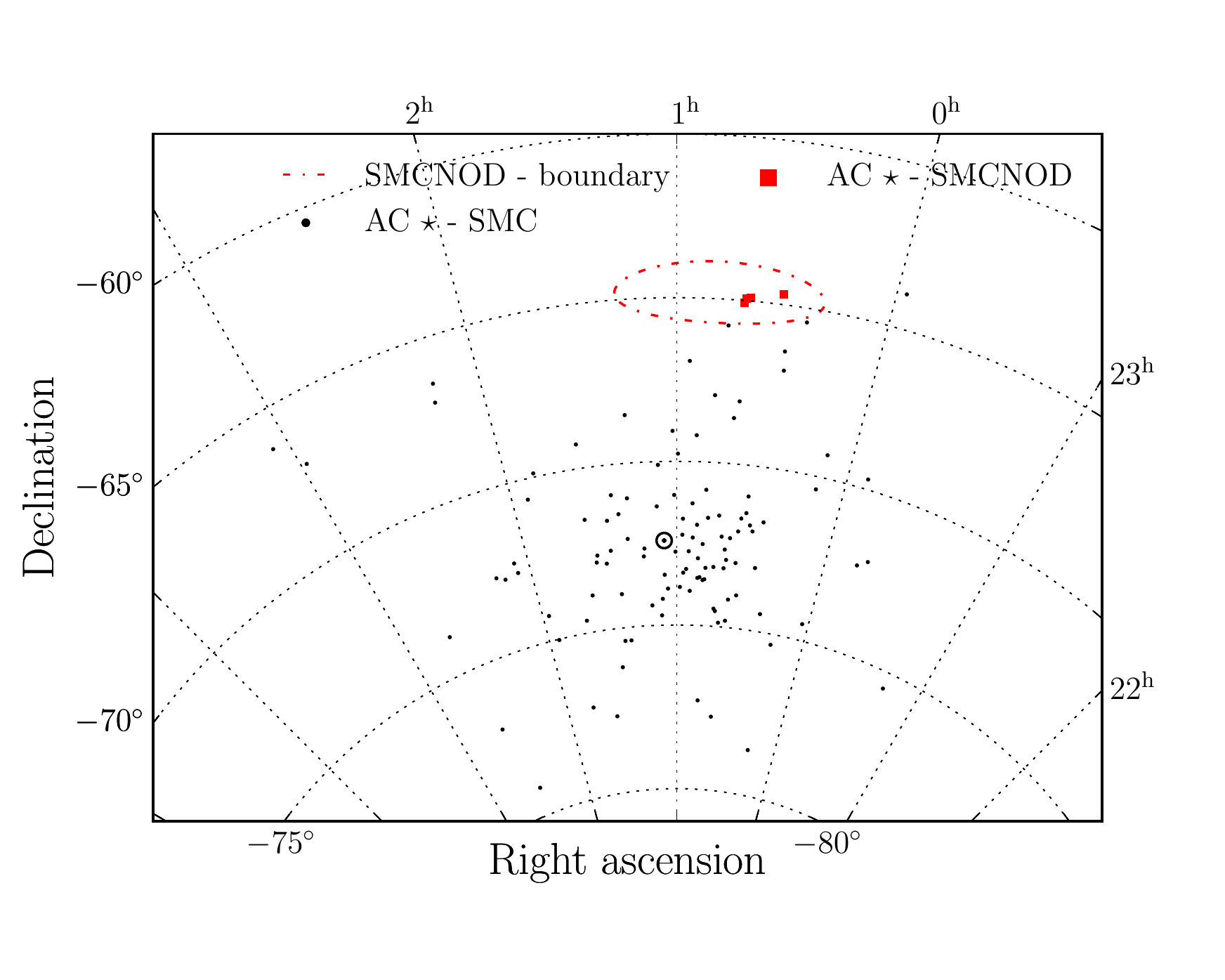}
\caption{Spatial distribution of ACs in the SMC. The red squares indicate ACs seen in projection against the SMCNOD location.}
\label{fig:SMCNOD-AC}
\end{figure}

The OGLE-IV photometry for the SMC contains 119 ACs \citep{Soszynski2015ACs}, of which approximately two-thirds pulsate in the fundamental mode while the rest are first-overtone pulsators. Fig.~\ref{fig:SMCNOD-AC} shows their spatial distribution. As noted by \cite{Soszynski2015ACs}, the radial distribution of ACs in the SMC is similar to that of its RR~Lyrae stars, but in the outskirts, there is an excess of ACs compared to RR~Lyrae stars \citep[see fig.~7 in][]{Soszynski2015ACs}. Four ACs are located at the same range of coordinates as the SMCNOD. If they do lie in the outskirts of the SMC, these four ACs would be somewhat unusual considering the paucity of ACs in other border regions of the SMC. In addition, their spatial concentration resembles  the densest regions of the SMC. To explore a possible membership of these four ACs in the SMCNOD we calculated their distances using the procedure from \cite{JD2016CCs} and the coefficients for ACs from \cite{Soszynski2015ACs}.

First, we calculated the reddening-insensitive Wesenheit magnitude \citep[$W_{I, V-I}$,][]{Madore1976} for each star:

\begin{equation}
W_{I, V-I} = I - 1.55 \cdot \left ( V - I \right ).
\label{eq:WesenKlas}
\end{equation}
$I$ and $V$ represent mean magnitudes in those passbands. For calculating errors we adopted the error of the zero point in the OGLE-IV photometry as $\sigma_{V,I}$ = 0.02\,mag \citep{JD2016CCs}. Subsequenty, we calculated the reference Wesenheit magnitude using the P-L relations from \cite{Soszynski2015ACs} for the dominant pulsation mode:

\begin{equation}
W_{\text{ref}} = a_{\text{SMC}} \cdot \text{log}P + b_{\text{SMC}},
\end{equation}
where $a_{\text{SMC}}=-2.85\pm 0.15$ and $b_{\text{SMC}}=17.01\pm 0.03$ for the fundamental mode and $a_{\text{SMC}}=-3.69\pm 0.15$ and $b_{\text{SMC}}=16.64\pm 0.05$ for the first overtone. The distances relative to the mean distance of the SMC were computed by

\begin{equation}
d = d_{\rm SMC} \cdot 10^{0.2 \cdot \left(W_{I, V-I} - W_{\rm ref - F/1O} \right) },
\label{eq:distanceACs}
\end{equation}
where for the distance to the SMC, we adopted the value from \cite{GrijsBono2015} of $d_{\rm SMC}$ = 61.94 $\pm$ 3.38\,kpc. The individual errors in the distances were calculated using error propagation. The basic properties of the four ACs that lie at the same coordinates as the SMCNOD are listed in Table~\ref{tab:SMCNOD-AC-table}. The distances for the remaining ACs can be found in the supplementary material online.

Fig.~\ref{fig:SMCNOD-AC-dist} shows a declination vs. distance plot of the ACs. The SMCNOD presumably lies at the same distance as the SMC \citep{Pieres2016SMCNOD}. The SMC has a significant depth in intermediate-age stars along the line of sight \citep[SMC disk 4.23\,$\pm$\,1.48\,kpc,][using red clump stars]{Subramanian2009}. Using the aformentioned depth, the scale height $h_{z}$ of the SMC is 1.97\,$\pm$\,0.69\,kpc \citep[using eq.~7 in][]{Haschke2012bb}. We did not derive the depth extent of the SMC using the ACs due to their low numbers in the SMC. Based on the analysis done by \citet{Pieres2016SMCNOD}, we assume that the SMCNOD has the same scale height as the SMC. By comparing the calculated distances and the SMCNOD scale height, four ACs, within the errors, are consistent with a position inside the SMCNOD.

\setlength{\tabcolsep}{5pt}
\begin{table}
\centering
\caption{Basic properties of ACs in the coordinate and distance range of the SMCNOD. Column 1 provides the star ID in the form OGLE-SMC-ACEP-ID, column 2 contains the pulsation mode, column 3 and 4 provide equatorial coordinates (J2000), and the last two columns contain distances with errors. The full table for all analyzed ACs in the SMC is available as supplementary material to this paper.}
\label{tab:SMCNOD-AC-table}
\begin{tabular}{lcccccc}
\hline
ID                & mode & R.A.\,[deg]      & Dec\,[deg]       & $d$\,[kpc]     & $\sigma_{d}$\,[kpc]  \\ \hline
016 & F    & 7.55325  & -64.69511 & 59.48 & 3.89 \\
027 & 1O   & 9.83200  & -64.90775 & 66.58 & 4.50 \\
031 & F    & 10.12183 & -64.93772 & 68.46 & 4.49 \\
034 & F    & 10.24825 & -65.06083 & 62.97 & 3.97 \\ \hline
\end{tabular}
\end{table}

\begin{figure}
\includegraphics[width=\columnwidth]{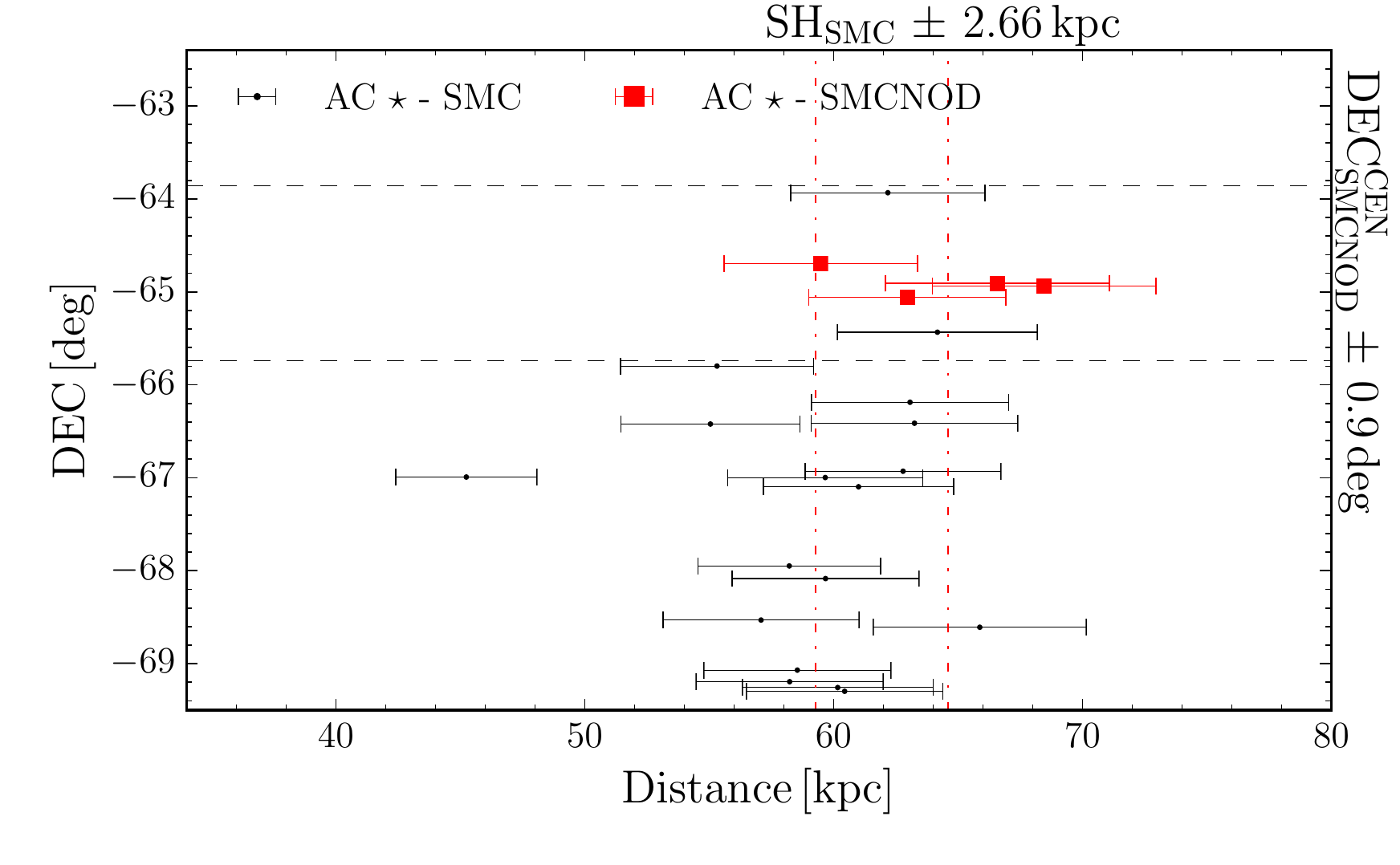}
\caption{Declination vs. distance for the ACs. The black dots represent ACs from the OGLE-IV photometry for the SMC and the red squares stand for ACs that presumably lie in the SMCNOD. The horizontal black dashed lines indicate lower and upper borders in declination for the SMCNOD. The red dash-dotted lines represent the assumed distance boundaries of the SMCNOD, based on the maximum scale height.}
\label{fig:SMCNOD-AC-dist}
\end{figure}

Apart from the distance estimation, ACs can constrain the ages of stellar populations in a given system. The assessment of the age of a given AC is done through its mass and absolute magnitude in the $V$ and $I$-band. To determine the $M_{V}$ and $M_{I}$ we used the extinction maps from \cite{Schlafly2011} (for the reddening $E(B-V)$ and extinction $A_{V}$) and equation 7 from \cite{Hamanowicz2016} to calculate the extinction in the $I$-band, $A_{I}$:

\begin{equation}
A_{I} = 1.969 \cdot E(B-V).
\end{equation} 
With the previously acquired distances we obtained estimates for the absolute magnitudes in both passbands. Following the procedure from \cite{Fiorentino2012} we determined masses of the ACs within the SMCNOD with the equations from \cite{Marconi2004}:

\begin{gather}
\text{log}M_{\text{F}} = -0.97 - 0.53 M_{I} - 1.55 \text{log}P + 1.44 \left ( M_{V} - M_{I} \right ) \\
\text{log}M_{\text{1O}} = -1.29 - 0.58 M_{I} - 1.79 \text{log}P + 1.54 \left ( M_{V} - M_{I} \right ),
\end{gather} 
where F and 1O represent the first overtone and fundamental mode, respectively. The resulting masses are in solar units. For the age estimate we used stellar evolutionary models from the BaSTI database\footnote{\url{http://albione.oa-teramo.inaf.it/}} \citep{Pietrinferni2004}. The results can be found in Table~\ref{tab:SMCNOD-AC-table-age}. We assumed non-canonical models with the commonly assumed mass loss parameter $\eta$ = 0.4 \citep{Girardi2000} and four different possible metallicities $Z$ = (0.0003, 0.0006, 0.001, 0.002). The broad metallicity range was selected based on the metallicity--age distribution \citep[see fig.~14 in][]{Cignoni2013} of the SMC. From this table, we see that the ACs inside the SMCNOD are most likely members of an intermediate-age population, which is in good agreement with the conclusions of \cite{Pieres2016SMCNOD}. We acknowledge that non-canonical models (with overshooting in the stellar interior) are more suitable for ages below 4\,Gyr during the hydrogen-burning phase. Thus we decided to recalculate ages also for canonical models (without overshooting) and compare them with ages calculated using non-canonical models. The difference between the two groups varies around 0.02\,Gyr and is therefore negligible at these ages.


\begin{table}
\centering
\caption{Table for calculated masses of the ACs that lie inside the SMCNOD. Column 1 lists star IDs in the same format as in Table~\ref{tab:SMCNOD-AC-table}, column 2 contains masses in solar units, and the subsequent columns 3 - 6 provide ages in Gyr for a given metallicity Z. We used evolutionary tracks with overshooting \citep[non-canonical models,][]{Pietrinferni2004} and we assumed a mass loss $\eta$ equal to 0.4.}
\label{tab:SMCNOD-AC-table-age}
\begin{tabular}{lccccc}
 & & Z=0.0003 & Z=0.0006 & Z=0.001 & Z=0.002 \\ \hline
ID & M [M$_{\odot}$] & Age\,[Gyr] & Age\,[Gyr] & Age\,[Gyr] & Age\,[Gyr] \\ \hline
016 & 1.261 & 2.31 & 2.36 & 2.44 & 2.63 \\
027 & 1.081 & 4.06 & 4.16 & 4.31 & 4.65 \\
031 & 1.261 & 2.31 & 2.36 & 2.44 & 2.63 \\
034 & 1.292 & 2.28 & 2.33 & 2.41 & 2.59 \\ \hline   
\end{tabular}
\end{table}

\subsection{RR~Lyrae stars}

RR~Lyrae stars are Population II stars, mostly radially pulsating variables, located on the horizontal branch of the Hertzsprung-Russell diagram. They can be used to study stellar pulsations and evolution and are unambiguous indicators of populations older than 9\,Gyr. RR~Lyrae stars are also excellent distance indicators and thus can be used to study the structure of the old population of the Milky Way and nearby galaxies. RR~Lyrae stars can be divided into three groups based on their pulsation modes. The vast majority of these stars are fundamental-mode (RRab -- F) or first-overtone (RRc -- 1O) pulsators. The least populous group consists of double-mode pulsators that pulsate in the fundamental and first-overtone simultaneously (RRd -- F+1O).

\begin{figure}
\includegraphics[width=\columnwidth]{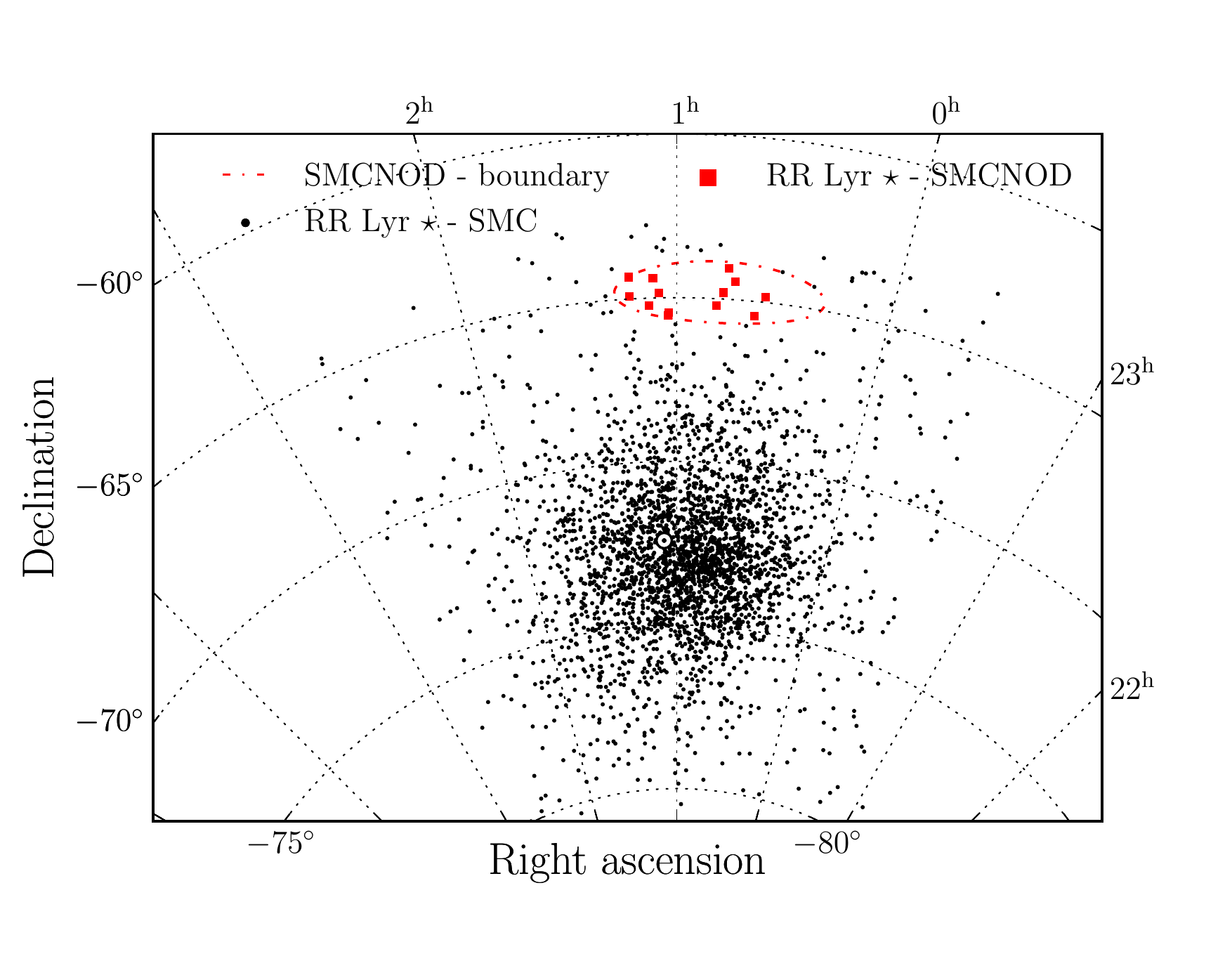}
\caption{Similar figure as the Fig.~\ref{fig:SMCNOD-EC} but for RR~Lyrae stars in the SMC.} 
\label{fig:SMCNOD-RRLyr}
\end{figure}

OGLE-IV contains over 45\,000 RR~Lyrae stars in the Magellanic system, of which 6\,572 lie in the SMC. Of the whole sample of RR~Lyraes in the SMC more than 5\,000 belong to the RRab type, over 800 are first overtone pulsators, and more than 600 are RRd-type variables. We decided to use only the fundamental-mode RR~Lyraes. Their spatial distribution is plotted in Fig.~\ref{fig:SMCNOD-RRLyr}. This figure demonstrates that RR~Lyrae stars show a roughly spheroidal distribution with a pronounced concentration towards the central regions of the SMC \citep[see also][]{Haschke2012bb,JD2016RRLyr}. Thirteen RR~Lyrae stars have celestial positions overlapping with the SMCNOD.  

In order to test whether these thirteen RR~Lyraes could plausibly be a part of the SMCNOD, we calculated distances of individual RR~Lyrae stars in the SMC. For the distance estimation, we used similar procedures as described in \cite{Skowron2016} and \cite{JD2016RRLyr}. For the distance calculation, we need to estimate the metallicity of the individual stars. We used equation 3 from \cite{Smolec2005} and excluded unsuitable RR~Lyrae stars from the analysis in the following manner. First, we removed stars that had no mean magnitude in the $V$-band and no value for the Fourier parameter $\varphi_{31}^{I}$. Second, we omitted stars with amplitudes of A$_{I}$ $< -5 \cdot \text{log}P - 1$ in order to remove possible blends. This cut removed 76 stars from the short-period end of the period-amplitude diagram. In addition, many RR~Lyrae stars exhibit a quasi-periodic modulation of the light curve known as the Blazhko effect \citep[for a review, see][]{Szabo2014}. It is assumed that almost 50\,\% of the fundamental-mode RR~Lyraes display this effect \citep[see][]{Prudil2017,Benko2010}. It is generally presumed that stars exhibiting the Blazhko effect are not suitable for determining their metallicities via a Fourier analysis of their light curves and thus subsequent distances. To remove stars that show this effect we used equation 2 from \cite{Prudil2017} and omitted RR~Lyraes that lie more than 4$\sigma$ below this fit. In the end, we were left with 2\,991 RRab-type pulsators. Subsequently, we fitted truncated Fourier-series to the $I$-band light-curves of the remaining stars, of the form

\begin{equation} \label{FourierSeries}
m\left ( t \right ) = A_{0}^{I} + \sum_{k=1}^{n=6} A_{k}^{I} \cdot \text{cos} \left (2\pi k \frac{HJD - M_{0}}{P} + \varphi_{k}^{I} \right ).
\end{equation}
In this equation the $A_{k}^{I}$ represent amplitudes, the heliocentric Julian date, $HJD$, is the time of the observation, $M_0$ is the time of brightness maximum, $P$ represents the pulsation period, and the $\varphi_{k}^{I}$ are the phases. The next step was to calculate the metallicity [Fe/H]$_{\text{S05}}$ using the relation from \cite{Smolec2005}:

\begin{equation} \label{FEH-SM}
{\rm[Fe/H]_{\text{S05}}} = b_{1} + b_{2} \cdot P + b_{3} \cdot \varphi_{\rm 31} + b_{4} \cdot A_{2}^{I},
\end{equation}
where the coefficients $b$ are $b_{1}=-6.125\pm 0.832$, $b_{2}=-4.795\pm 0.285$, $b_{3}=1.181\pm 0.113$, and $b_{4}=7.876 \pm 1.776$. The errors in the metallicity derived from Eq.~\ref{FEH-SM} were estimated using a Monte Carlo error simulation. Using Equation \ref{FEH-SM} we obtained metallicities on the \cite{Jurcsik1995} scale. To calculate distances, we first transformed this metallicity to the \cite{Carretta2009} scale through the equation 3 from \cite{Kapakos2011}:

\begin{equation}
{\rm[Fe/H]_{\text{C09}}} = 1.001 \cdot {\rm[Fe/H]_{\text{J95}}} - 0.112.
\end{equation}
With the resulting metallicities, we could then proceed with the distance estimation in the following manner. First, we calculated the Wesenheit magnitude for an individual RR~Lyrae star using Equation \ref{eq:WesenKlas}. Afterwards, we used an equation from \cite{Braga2015} to calculate the absolute Wesenheit magnitude:

\begin{equation}
W_{I, V-I}^{abs} = a_{1} + a_{2} \cdot {\rm log}P + a_{3} \cdot \left ( {\rm [Fe/H]_{C09}} + 0.04 \right ),
\label{WesenAbsMag}
\end{equation}
where the parameters $a$ are $a_{1}=-1.039\pm 0.007$, $a_{2}=-2.524\pm 0.021$ and $a_{3}=0.147\pm 0.04$. Subsequently, using the distance modulus we got the distance in kpc:

\begin{equation} \label{distanceRRLyr}
d = 10^{\left (W_{I, V-I}-W_{I, V-I}^{abs} + 5  \right )\cdot 0.2}.
\end{equation}
The errors for individual distances were calculated using error propagation, where we assumed $\sigma_{V,I} = 0.02$\,mag \citep{JD2016RRLyr} for the uncertainties in the OGLE-IV passbands $V$ and $I$. The basic properties of the RR~Lyrae stars that coincide in projection with the SMCNOD can be found in Table~\ref{tab:SMCNOD-RRL-table}. 

\begin{figure}
\includegraphics[width=\columnwidth]{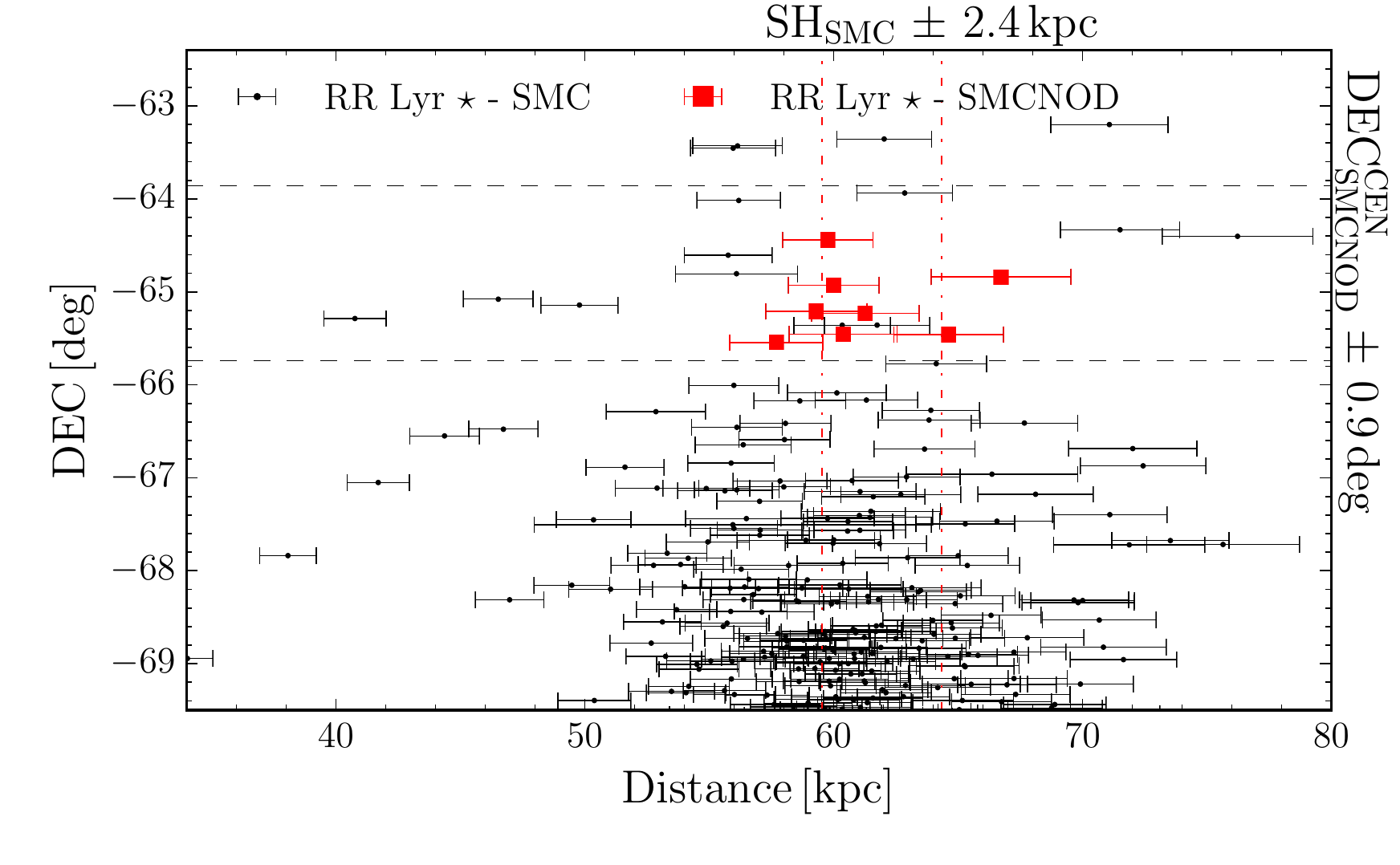}
\caption{Same as Fig.~\ref{fig:SMCNOD-AC-dist}, but for RR~Lyrae stars. The red squares stand for RR~Lyraes that probably belong the SMCNOD. The black dots represent the remaining RR~Lyraes in the SMC. The red dash-dotted lines represent assumed distance boundaries for the SMCNOD, based on the maximum scale height. In addition, the dashed horizontal lines represent the adopted boundary in declination for the SMCNOD.}
\label{fig:SMCNOD-RRL-dist}
\end{figure}

Fig.~\ref{fig:SMCNOD-RRL-dist} shows the declination vs. distance for individual stars. To constrain the RR Lyraes based on their possible association with the SMCNOD we assumed the scale height of the old population of the SMC to be 2.0\,$\pm$\,0.4\,kpc \citep{Haschke2012bb}, thus within a possible range up to 2.4\,kpc. Overall, out of 13 RR~Lyrae stars that share the coordinates with the projected location of the SMCNOD, 8 are consistent with the assumed scale height for the SMCNOD within the errors. This result is based on the assumption that SMCNOD shares the same scale height as the SMC.

\setlength{\tabcolsep}{2pt}
\begin{table} 
\centering
\caption{Properties of RR~Lyrae stars that lie at the same coordinates as the SMCNOD. Column 1 contains the ID of a star in the form OGLE-SMC-RRLYR-ID, columns 2 and 3 provide right ascension and declination (J2000), columns 4 and 5 contain distances with errors, similar to columns 6 and 7 that contain metallicities and their errors. The full table for all analyzed RR~Lyrae stars in the SMC is available as supplementary material to this paper.}
\label{tab:SMCNOD-RRL-table}
\begin{tabular}{lcccccc}
\hline 
ID                  & R.A.\,[deg]       & Dec\,[deg]       & $d$\,[kpc]     & $\sigma_{d}$\,[kpc]    & [Fe/H]    & $\sigma_{\rm [Fe/H]}$  \\ \hline
3542 & 8.78842  & -64.83892 & 66.73 & 2.82 & -1.98 & 0.45 \\
3633 & 9.43142  & -65.45781 & 60.39 & 2.16 & -1.72 & 0.31 \\
3851 & 10.96179 & -64.44103 & 59.76 & 1.82 & -1.70 & 0.14 \\
3941 & 11.47646 & -64.06644 & 5.85  & 0.17 & -1.29 & 0.04  \\
3973 & 11.73700    & -64.80486 & 56.10 & 2.45 & -1.27 & 0.49 \\
4054 & 12.19496 & -65.20972 & 59.31 & 2.04 & -0.78 & 0.29 \\
4566 & 15.55129 & -65.45931 & 64.63  & 2.20 & -2.18 & 0.25 \\
4581 & 15.63971 & -65.54542 & 57.70 & 1.87  & -1.59 & 0.22 \\
4685 & 16.24354 & -64.84886 & 17.29 & 0.51 & -1.72 & 0.07 \\
4738 & 16.63004 & -64.40250  & 76.23 & 3.03 & -2.50  & 0.39 \\
4777 & 16.94008 & -65.23344 & 61.27 & 2.16 & -2.05 & 0.29 \\
4960 & 18.25642 & -64.33211 & 71.51 & 2.39 & -2.27 & 0.23 \\
4965 & 18.30121 & -64.92725  & 60.01 & 1.82 & -1.56 & 0.14 \\ \hline
\end{tabular}
\end{table}

\section{Analysis of the spatial distribution} \label{sec:FieldStaAnalysis}

In this section, we explore whether the four ACs and eight RR~Lyrae stars whose positions and distances coincide with the SMCNOD stand out in comparison with the overall stellar densities in the SMC outskirts. We created a set of 14 contiguous ellipses in equatorial coordinate space with the same size as the SMCNOD in the previous plots (see Figs.~\ref{fig:SMCNOD-EC}, \ref{fig:SMCNOD-AC}, \ref{fig:SMCNOD-RRLyr} for reference). These ellipses were distributed around the SMCNOD in the outskirts of the OGLE-IV photometry and include the northern boundaries of the SMC. Their minor and major axes, at the distance of the SMC, are 1.02 and 7.98\,kpc, respectively.

\subsection{Anomalous Cepheids} \label{subsec:FSA-AnomCep}

\begin{figure}
\includegraphics[width=\columnwidth]{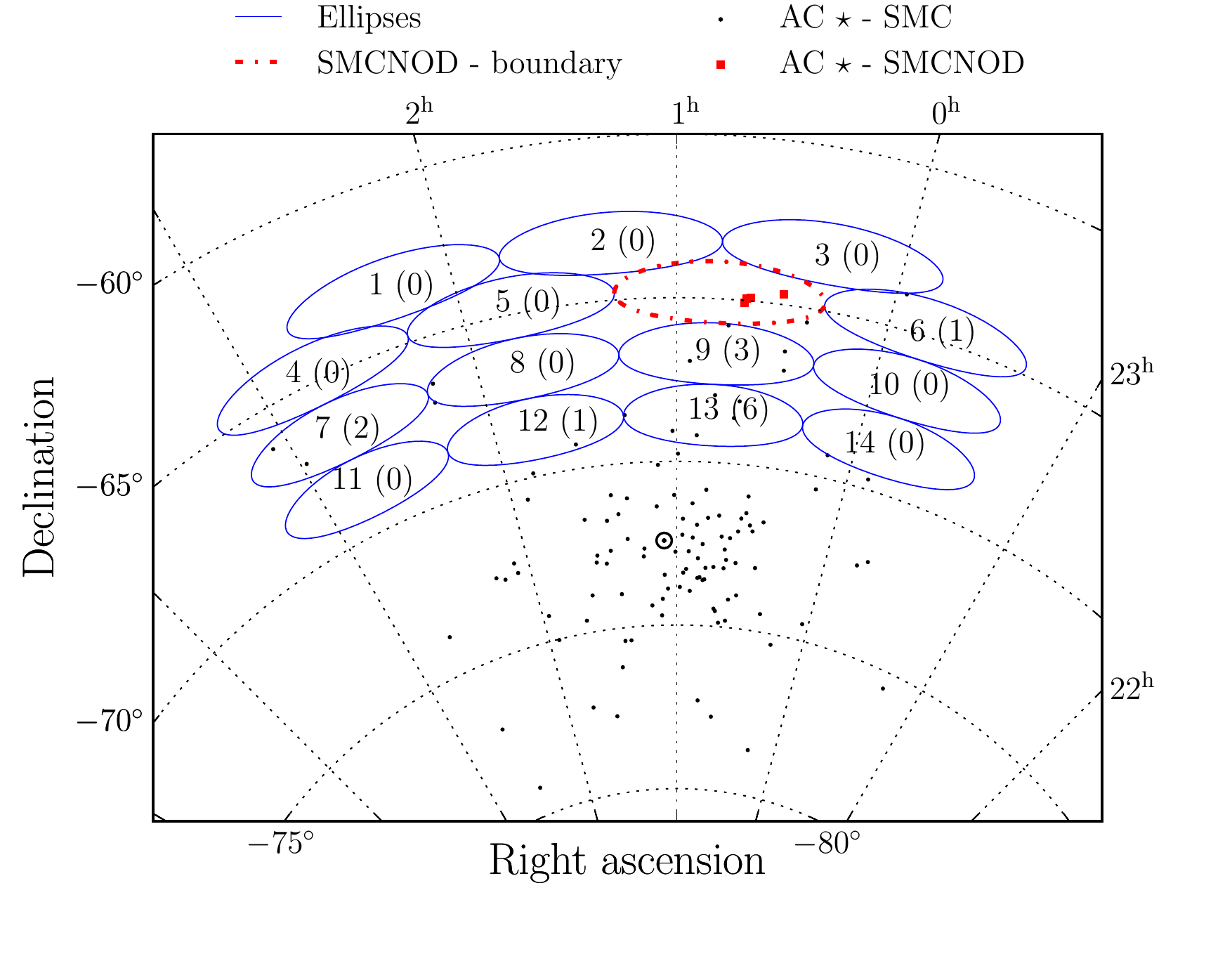} %
\caption{Field star analysis for the ACs in the SMC. The panel shows the spatial distribution of ACs in the SMC with the red dash-dotted ellipse marking the position of the SMCNOD and blue ellipses representing the test fields. The numbers inside the ellipses stand for the designation of the given ellipse with the number of ACs within the scale height (number in brackets). The ACs that lie inside the SMCNOD are represented by red squares. Black dots stand for ACs in the SMC. The Sun symbol denotes the apparent kinematic center of the SMC \citep{Stanimirovic2004}.}
\label{fig:SMCNOD-FSA-AC}
\end{figure}

First, we looked at the ACs and their distribution in the constructed ellipses (see the Fig.~\ref{fig:SMCNOD-FSA-AC}). The red dash-dotted ellipse represents the SMCNOD and the blue ellipses represent the area considered in addition. The meaning of the points is the same as in Fig.~\ref{fig:SMCNOD-AC}. From this plot, we clearly see that only some of the surrounding ellipses contain AC stars (ellipses 6, 7, 9, 12 and 13). In addition, the ellipses 9 and 13 contain the majority of the ACs that fell into the outlined boundaries (nine out of thirteen). 



\begin{figure}
\includegraphics[width=\columnwidth]{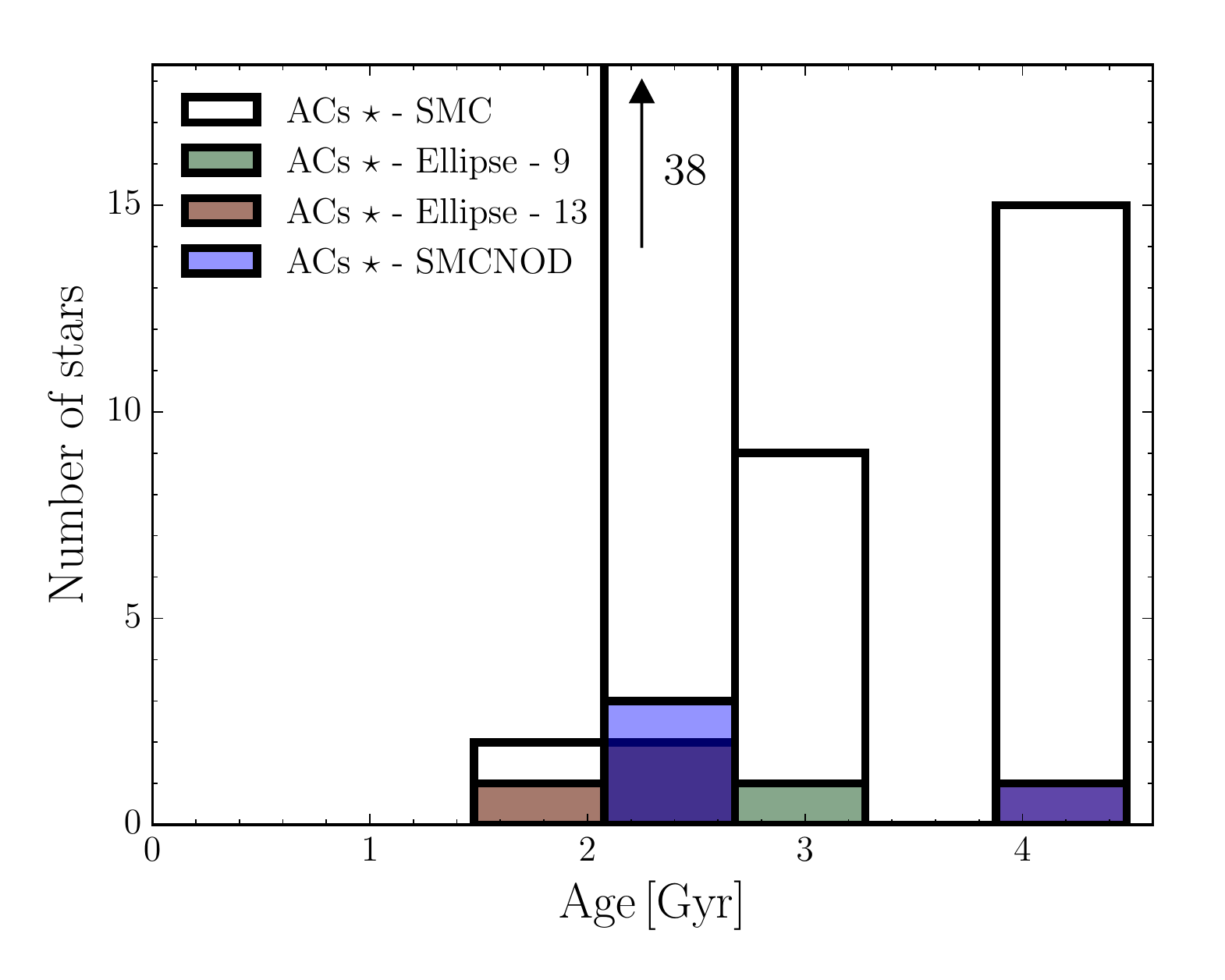} %
\caption{Distribution of ages for ACs in the SMC (transparent), Ellipse 9 (green columns), 13 (red columns) and SMCNOD (blue columns). To increase clarity, we decreased the vertical axis range and marked the best populated bin range with an arrow and the number of the stars included.} 
\label{fig:SMCNOD-AGE-AC}
\end{figure}

Therefore, the ACs found at the coordinates of the SMCNOD together with ellipses 9 and 13 appear to constitute an extension of the distribution of the ACs from the SMC. In addition, four ACs from the SMCNOD seem to form a tight group in coordinate space. Such an overdensity of ACs in equatorial coordinates in the SMC outskirts stands out as compared with other regions of the SMC outskirts. Furthermore, applying the procedure from Section \ref{sec:AC-dista-klasika} we estimated ages and masses for the ACs within the ellipses 9 and 13 and compared them with the values for the SMCNOD. We found that the ages for the ACs in these ellipses are approximately in the same range as for the SMCNOD ACs (around 2.8\,Gyr for Z=0.001). Therefore, the ACs in ellipses 9 and 13 most likely also belong to the intermediate-age population (see Fig.~\ref{fig:SMCNOD-AGE-AC}). 

\subsection{RR~Lyrae stars}

\begin{figure}
\includegraphics[width=\columnwidth]{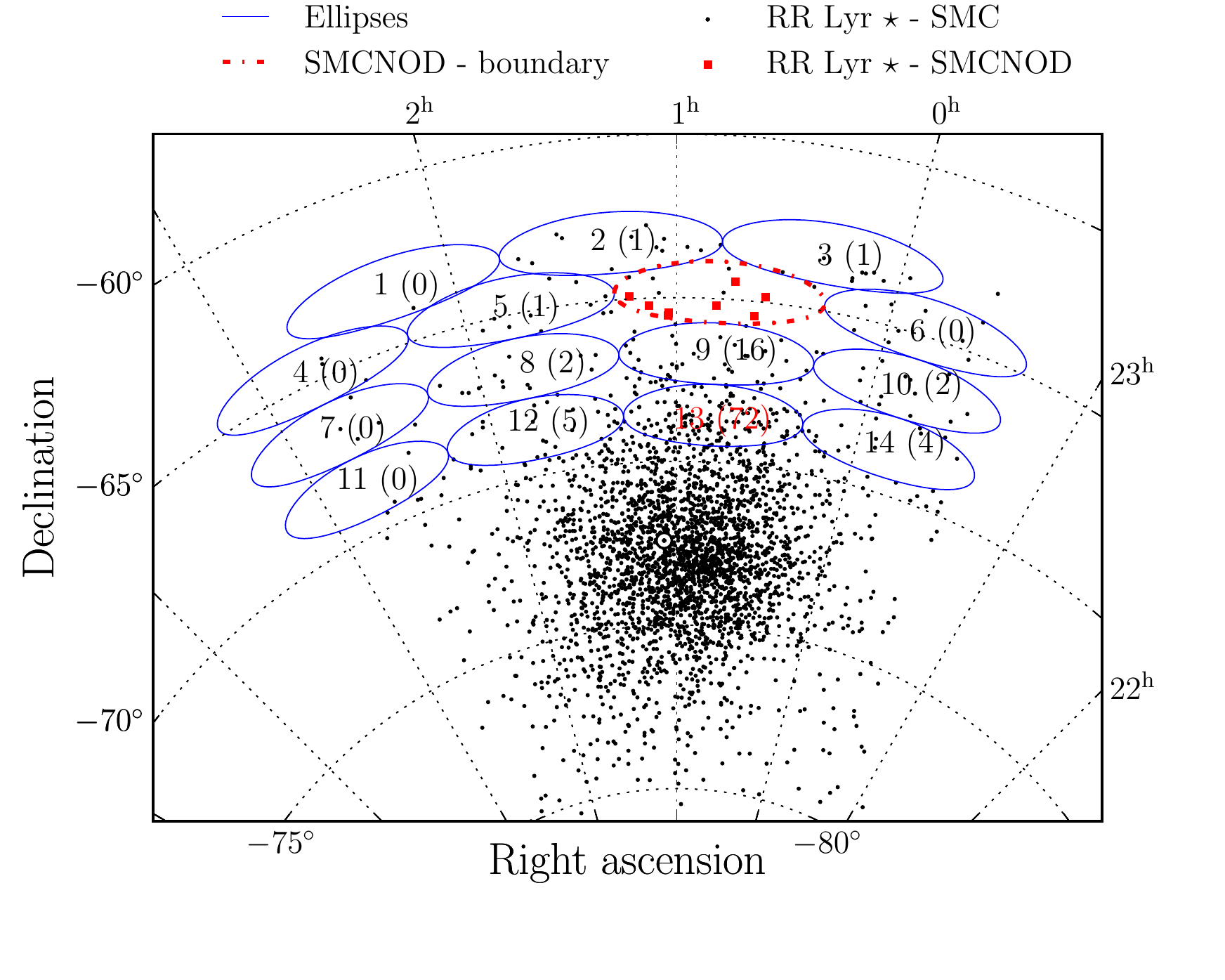} %
\caption{Same as in Fig.~\ref{fig:SMCNOD-FSA-AC} but for RR~Lyrae stars of type ab.} 
\label{fig:SMCNOD-FSA-RRLYR}
\end{figure}

In Fig.~\ref{fig:SMCNOD-FSA-RRLYR} we analysed the projected spatial distribution of the RR~Lyrae stars by subdividing the northern outer regions of the SMC into elliptical fields of the approximate size of the SMCNOD. The same symbols as in Fig.~\ref{fig:SMCNOD-FSA-AC} are used. We see an excess of RR~Lyraes in the ellipses 9 and 13 that are located between the central region of the SMC and the SMCNOD. The SMCNOD looks almost like an extension of this higher-density region. In the ellipses 9 and 13, we also see a large number of RR~Lyraes that share the same scale height as the main body of the SMC. Of the total number of 31 RR~Lyrae stars in ellipse 9, 16 stars lie within one scale height of the SMC. For ellipse 13, 72 out of total number 123 RR~Lyraes share the distance range of one SMC scale height, which is something we would expect based on the density profile for RR~Lyrae stars in the SMC (see \S~\ref{subsec:CCsDenProf}).

\begin{figure*} 
\includegraphics[width=510pt]{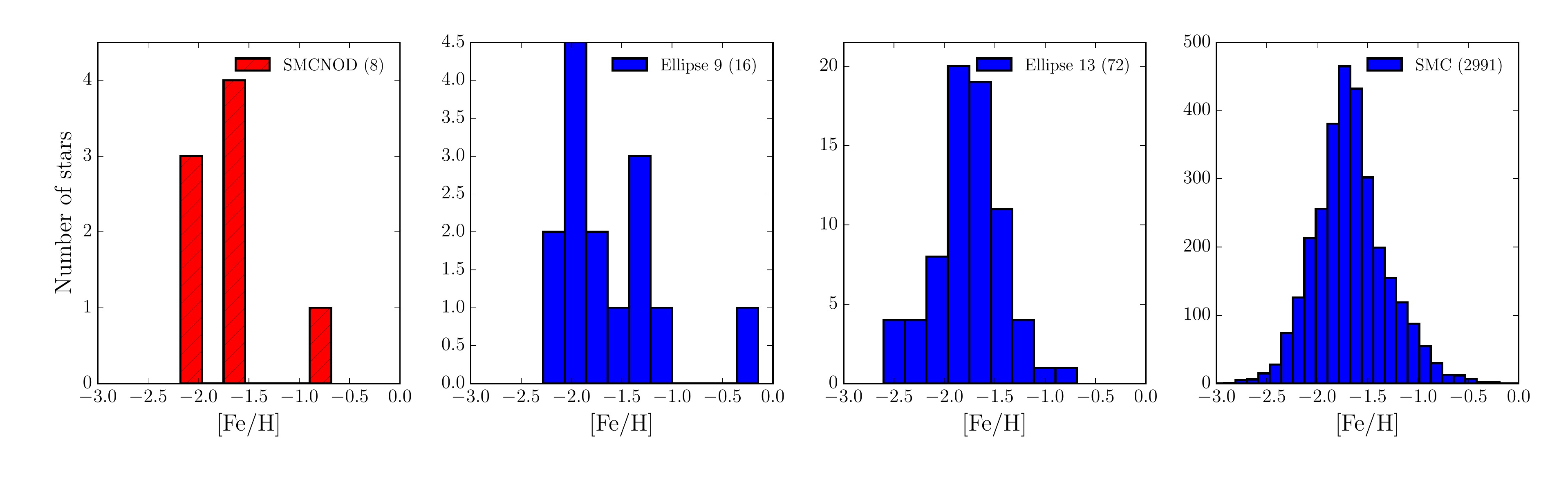} 
\caption{Metallicity distribution for the SMCNOD, selected ellipses, and the SMC overall. The vertical axis depicts the number of stars and the horizontal axis represents metallicity \citep[on the][scale]{Jurcsik1995} with the same range for all distributions. The numbers in brackets represent the total number of RR~Lyraes shown.}
\label{fig:SMCNOD-FSA-RRLYR-METAL}
\end{figure*}

The metallicity distribution (on the \cite{Jurcsik1995} scale) is shown in Fig.~\ref{fig:SMCNOD-FSA-RRLYR-METAL} for ellipses close to the SMCNOD and with a large number of RR~Lyrae stars. Only stars that fell into a distance range of one scale height of the SMC (and SMCNOD, respectively) were included in this plot. Other ellipses were too distant in equatorial coordinates (ellipse 12 - containing five RR~Lyraes) or contain only a few RR~Lyrae at a set scale height range (ellipse 8 - containing two RR~Lyraes) for the SMC. The median values of the metallicities are $\text{[Fe/H]}_{\rm SMCNOD}=-1.71\pm0.21$\,dex, $\text{[Fe/H]}_{\rm Ellipse-9}=-1.79\pm0.28$\,dex and $\text{[Fe/H]}_{\rm Ellipse-13}=-1.75\pm0.20$\,dex. The number of RR~Lyrae stars in each case were 8, 16, and 72, respectively. The median metallicity calculated for RR~Lyraes in the SMC in our study is $-1.69\pm0.21$\,dex. For comparison, \citet{Skowron2016} found the metallicity of the RR~Lyrae population of the SMC to be $-1.77\pm0.48$\,dex (on the \citet{Jurcsik1995} scale, using 3\,560 fundamental mode RR~Lyraes) or $-1.70$\,dex with an intrinsic spread of the metallicity distribution function of $0.27$\,dex on the \citet{Zinn1984} scale \citep{Haschke2012c}. It is clearly seen from the metallicities that the ellipses 9 and 13 within the same R.A. range share similar metallicities with the SMCNOD and the SMC as a whole. Only ellipse 8, with different R.A. ranges, has a different median metallicity ($\text{[Fe/H]}_{\rm Ellipse-8}=-2.79\pm0.53$\,dex) than the other plotted ellipses. We acknowledge that within the errors, the metallicities are similar. In addition, we note that low number statistics might have affected our results for ellipse 8, since it contains only 2 RR~Lyraes. Furthermore, the very low metallicity for ellipse 8 should be taken with a grain of salt, since the empirical relation for the metallicity of RR~Lyraes has been calibrated only up to -2.1\,dex.

\subsubsection{RR~Lyrae subtypes}

\begin{figure*} 
\includegraphics[width=510pt]{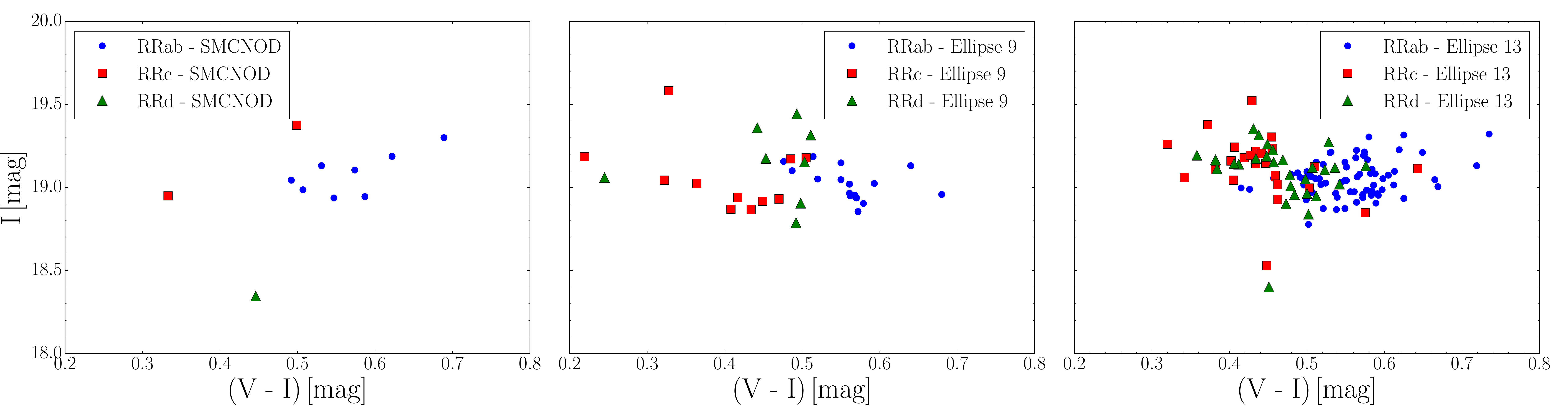} 
\caption{From left to right: Color-magnitude diagrams for SMCNOD, ellipse 9, and 13, respectively. The blue points represent the fundamental mode pulsators, red squares stand for the first-overtone variables and green triangles represent RRd type stars.}
\label{fig:SMC-FSA-RRLyr-CMD}
\end{figure*}

Up to this point, we only considered RR~Lyraes that pulsate in the fundamental mode. However, the first-overtone and double-mode pulsators also occupy the northern outskirts of the SMC. Distance estimation for RRc/RRd type stars can be onerous, therefore we decided to base our reasoning strictly on their position in the CMD (see Fig.~\ref{fig:SMC-FSA-RRLyr-CMD}) for a given ellipse. The blue points represent fundamental mode pulsators, red points stand for the first-overtone variables, and green triangles correspond to RRd type stars.

Based on equatorial coordinates and location in the CMD, the SMCNOD contains probably two first-overtone RR~Lyrae out of four potential candidates. On the other hand, in the ellipses below the SMCNOD (9 and 13 thus closer to the SMC main body) the situation is somewhat different. The aforementioned ellipses contain in total 17 and 26 RRc type stars, respectively. 

Out of these, approximately eleven first-overtone pulsators in ellipse 9 and 25 in ellipse 13 lie in a similar magnitude region in the CMD as the fundamental mode pulsators found at the distance range of the SMC. Therefore, the most populated ellipses with the RRab pulsators in the SMC outskirts contain a quite significant number of RRc-type stars. In general in the SMC for each RRc-type star, we can find more than 6 fundamental-mode pulsators (5\,105 RRab-type stars and 801 RRc-type stars). Therefore the number of first-overtone pulsators in the SMCNOD and in ellipse 13 roughly corresponds to the expected ratio of RRc-type stars. On the other hand, the nearby ellipse 9 contains a higher number of first-overtone variables.

Double-mode pulsators can also be found in the coordinate range of the SMCNOD and in ellipses 9 and 13. Based on its position in the CMD, the SMCNOD presumably contains only one RRd-type star. The aforementioned ellipses contain a higher number of double-mode pulsators. Ellipse 9 contains 8 RRd-type stars. Based on the CMD all lie within the scale height of the SMC. For ellipse 13 we found 28 RRd type variables to share the position in the coordinate space and all of them seem to lie within the scale height of the SMC. Similar to the first-overtone pulsators, double-mode variables are also abundant in ellipse 9. Ellipse 13 and the SMCNOD, on the other hand, contain on average the expected ratio of double-mode pulsators in comparison with the overall ratio for the SMC.

\section{Detection limitations of the SMCNOD} \label{sec:DLS}

In this section, we explore the possibility whether variable stars that were found inside the SMCNOD truly create an overdensity or are just a part of the expected density profile of the SMC outskirts. 

\subsection{Cartesian coordinate system and density profile} \label{subsec:CCsDenProf}

\begin{figure*}
\includegraphics[width=504pt]{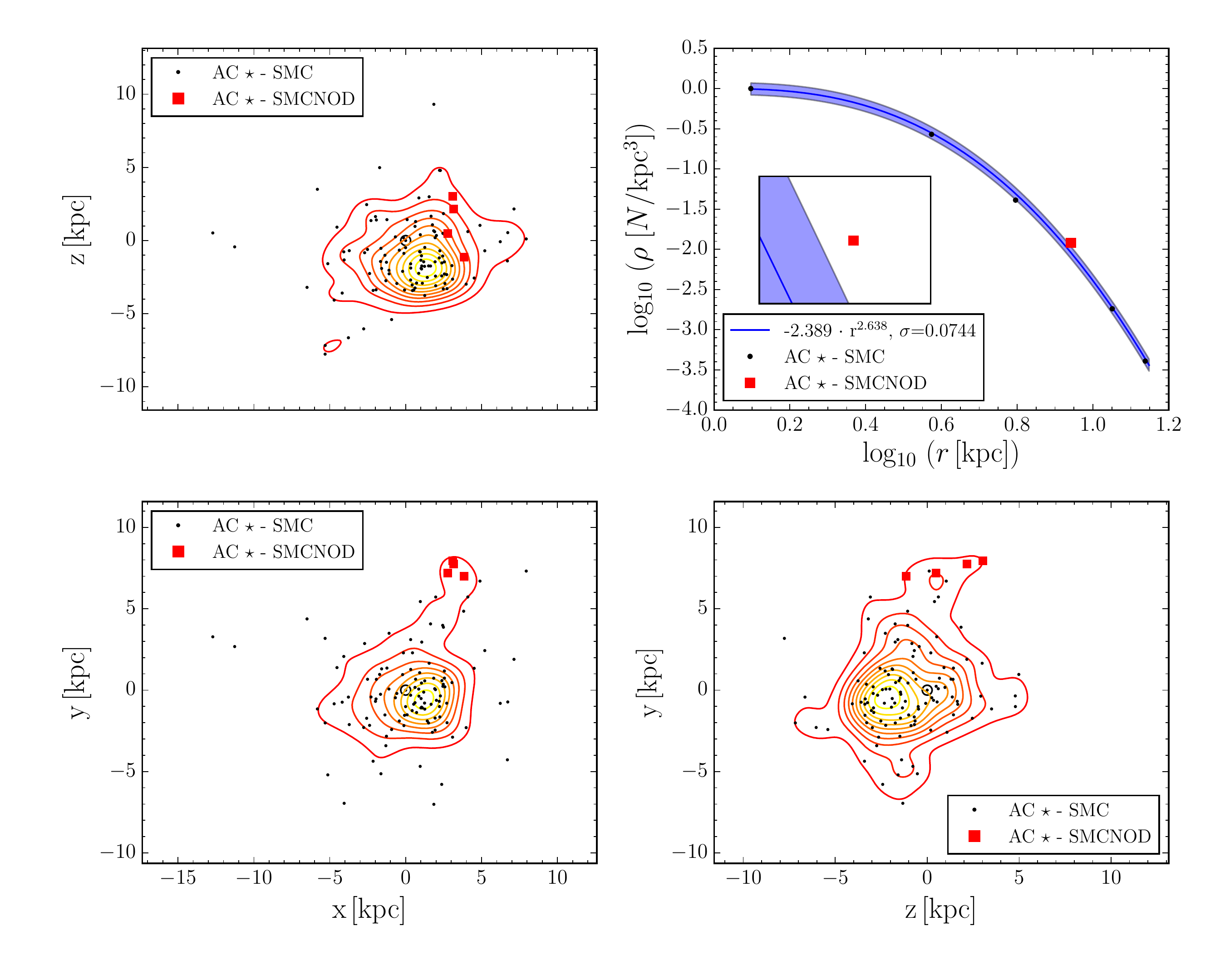} 
\caption{The top left and the two bottom panels display ACs (black dots) in the Cartesian coordinate system centered on the SMC's dynamical center. The red squares represent ACs found at the equatorial coordinates of the SMCNOD. The Sun symbol denotes the apparent kinematic center of the SMCNOD. In addition, we overplotted the spatial distribution of ACs using kernel density estimation with ten contour levels. The top right panel displays the density profile for ACs calculated based on their apparent kinematic center and stretching towards the outskirts of the SMC (with bin size 2.5\,kpc). The black points represent bins for ACs and the red square (at log$_{\rm 10}r=0.942$\,kpc) stands for four ACs found in the SMCNOD.}
\label{fig:SMCNOD-DLSxyz-AC}
\end{figure*}

\begin{figure*}
\includegraphics[width=504pt]{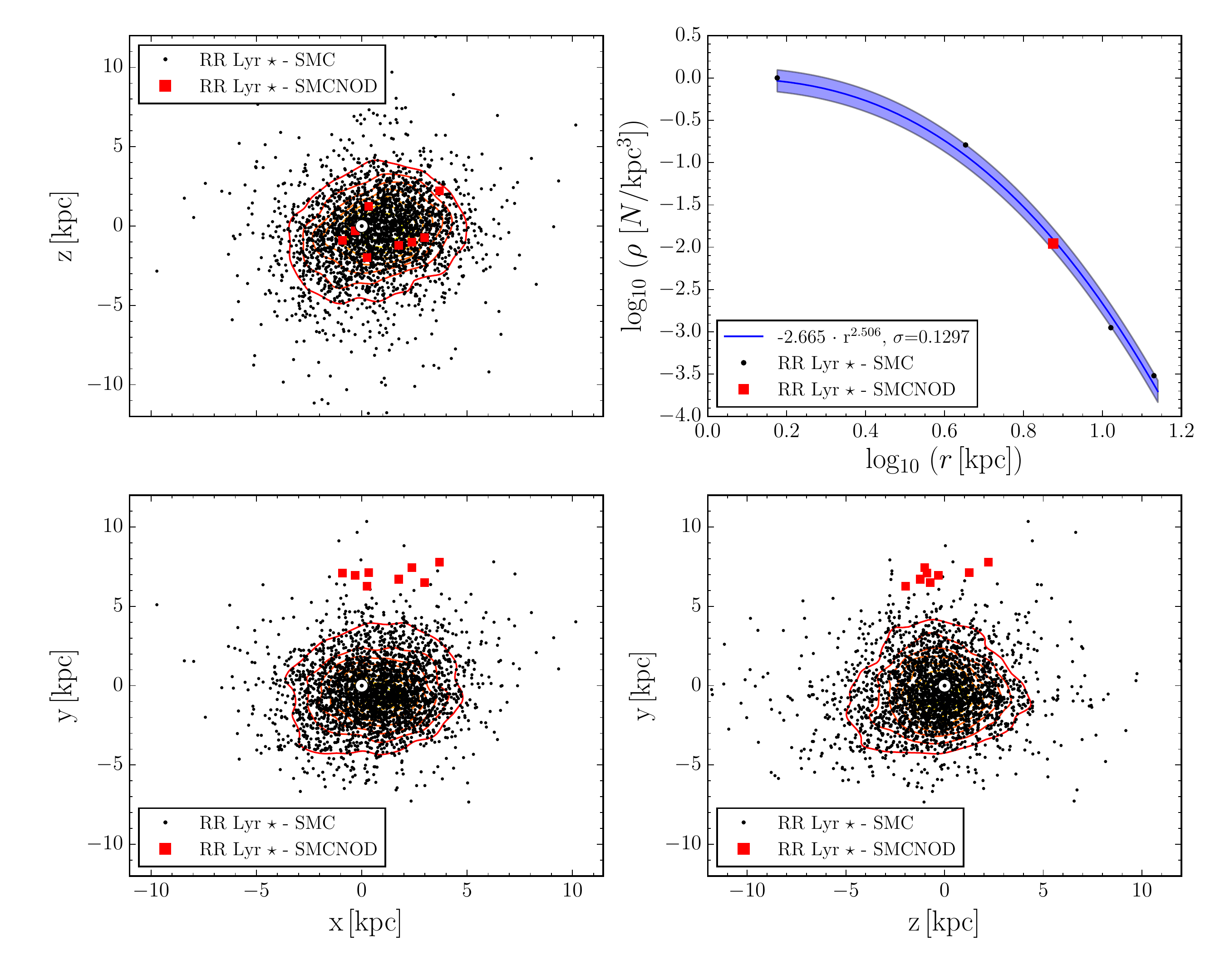} 
\caption{Similar to Fig.~\ref{fig:SMCNOD-DLSxyz-AC}, but for RR~Lyrae stars. The density profile in the top right corner was constructed by using annuli with a width of 3\,kpc. In addition, the red square in the upper right panel (at log$_{\rm 10}r=0.875$\,kpc) represents RR~Lyrae stars in the SMCNOD based on their radius $r$ with respect to the kinematic center of the SMC.}
\label{fig:SMCNOD-DLSxyz-RRLyr}
\end{figure*}

In order to test whether the SMCNOD can be detected using variable stars, we transformed equatorial coordinates and distances for ACs and RR Lyrae stars into a three-dimensional Cartesian coordinate system. We used the equations 12, 13, and 14 from \cite{JD2016CCs} in the following form:

\small
\begin{equation} 
x = -d \cdot \rm cos \left ( Dec \right ) sin \left ( R.A. - R.A._{\rm cen} \right ) 
\end{equation}
\begin{equation} 
\begin{split}
& y = d \cdot \rm \left ( sin \left ( Dec \right ) cos \left (Dec_{\rm cen}  \right ) \right ) -  & \\
& \hspace{1cm} d \cdot \rm cos \left ( Dec \right ) sin \left (Dec_{\rm cen}  \right ) \rm cos \left ( R.A. - R.A._{\rm cen} \right )   &
\end{split}
\end{equation}
\begin{equation} 
\begin{split}
& z = \left (d - d_{\rm SMC} \right )\cdot \rm \left ( cos \left ( Dec \right ) cos \left (Dec_{\rm cen}  \right ) \rm cos \left ( R.A. - R.A._{\rm cen} \right )\right) + \\
& \hspace{1cm} \left (d - d_{\rm SMC} \right )\cdot \rm sin \left ( Dec \right ) sin \left (Dec_{\rm cen}  \right )   ,
\end{split}
\end{equation}
\normalsize
where R.A.$_{\rm cen}$ and Dec$_{\rm cen}$ denote the coordinates of the apparent kinematic center of the SMC \citep{Stanimirovic2004}. The results of this transformation for ACs are depicted in Fig.~\ref{fig:SMCNOD-DLSxyz-AC}. The top left and the two bottom panels of Fig.~\ref{fig:SMCNOD-DLSxyz-AC} show ACs in the SMC in a 3D Cartesian coordinate system overplotted with the kernel density estimate. In the $x$ vs. $y$ plane, ACs found in the SMCNOD cluster in a similar fashion as in Fig.~\ref{fig:SMCNOD-AC}, while in $x$ vs. $z$ and $y$ vs. $z$ they follow a narrow sequence. In addition, this tail sequence seems to be elongated towards the larger distances. This effect is also observed in RR Lyrae stars (see Fig.~\ref{fig:SMCNOD-DLSxyz-RRLyr}). This effect can be related to the interaction between the Magellanic clouds since we a see similarly elongated distribution of the CCs in the SMC. We note that accounting for errors this trend could disappear. The panel on the top right displays the density profile of the ACs in the SMC, which was constructed in the following way: 

First, we utilized Cartesian coordinates and calculated an elliptical radius for the elongated shape of the SMC using the median axis ratio $1:1.10:2.13$ from \cite{JD2016RRLyr} and the following equation:

\begin{equation}
r = \sqrt{x^{2} + y^{2}\text{p}^{-2} + z^{2}\text{q}^{-2}},
\end{equation}
where p and q represent the ratios between $y$-$x$ ($1.10/1.0$) and $z$-$x$ ($2.13/1.0$), respectively \citep{JD2016RRLyr}. The radius $r$ then served as a scale for the density profile, where we counted the number of stars from the center of the SMC towards the outskirts in 3D shell-like bins. The bin size of the individual ranges in the density profile was 2.5\,kpc in order to encompass all four ACs found in the SMCNOD. From this plot, we see that the density of ACs at the galactocentric distance of the SMCNOD does slightly deviate from the power-law (blue line) fitted to the density profile of the ACs. This anomaly lies more than one $\sigma$ above the density profile. We note that the lower number of ACs in the northern outskirts of the SMC may have affected our results.

In Fig.~\ref{fig:SMCNOD-DLSxyz-RRLyr} the distribution of the SMC RR~Lyrae stars is plotted in the Cartesian coordinate system (top left and bottom panels). The top right panel shows the density profile of their distribution. From the Cartesian distribution, we see that RR~Lyrae stars are numerous in the outskirts of the SMC. They do not seem to form any noticeable overdensity at the position of the SMCNOD (see contours for kernel density estimation). A similar situation can be seen also for the density profile of RR~Lyrae stars. At the position of the SMCNOD, we do not observe any anomaly that would stand out. It is important to note that the bin sizes for ACs and RR~Lyrae stars were selected to comprise as many stars associated with the SMCNOD as possible. Therefore, our density profiles are slightly contaminated with stars occupying the same bin range (2.5\,kpc and 3\,kpc for ACs and RR~Lyraes, respectively). Thus, despite this contamination in the density profile for old population RR~Lyrae stars, we do not observe a statistically significant deviation that would stand out in the outskirts of the SMC.

\subsection{Probability of finding stars in the SMCNOD} \label{subsec:probSMCNOD}

For this analysis, we assumed that the distribution in the Cartesian coordinate system ($x$, $y$, and $z$) of ACs and RR~Lyraes in the SMC is Gaussian. Then, we simply integrated Gaussian functions for the Cartesian coordinates with respect to the boundaries set by variables inside the SMCNOD.

\begin{figure*}
\includegraphics[width=504pt]{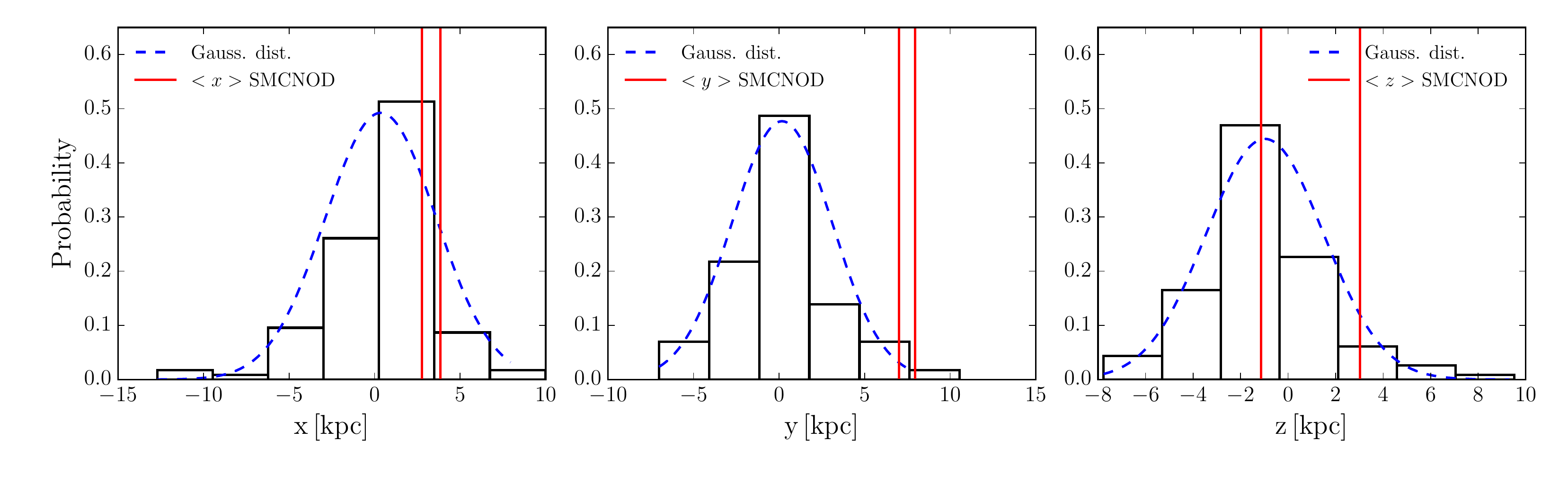} \\%
\includegraphics[width=504pt]{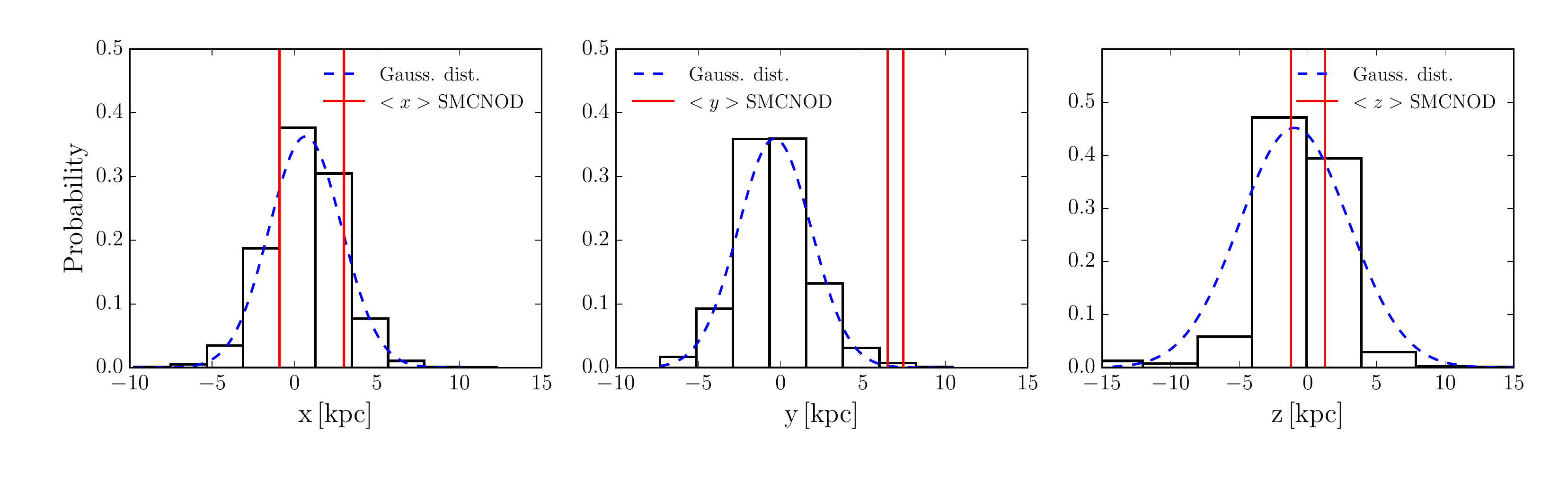} %
\caption{The distribution of Cartesian coordinates for ACs (top three panels) and RR~Lyraes (bottom three panels). The dashed blue lines represent a Gaussian distribution and the red vertical lines mark the integration ranges of a given coordinate for variables found in the SMCNOD.}
\label{fig:SMCNOD-PRO-AC-RRL}
\end{figure*}

First, we analyzed ACs, using the positions of the ACs in the SMCNOD as boundary conditions for the integration of the three distributions. In the top panel of Fig.~\ref{fig:SMCNOD-PRO-AC-RRL} we see the distributions for $x$, $y$, and $z$ coordinates overplotted with the Gaussian distribution (blue dashed lines). The red vertical lines denote the average values of ACs found in the SMCNOD for a given dimension. The probabilities calculated for each coordinate were multiplied with the number of ACs in the SMC and divided by the number of ACs in the SMCNOD. The resultant probability of finding four ACs at the position of the SMCNOD is less than 1.0\,\%.

To estimate the probability of finding eight RR~Lyrae stars at the position of the SMCNOD we proceeded in the same way as in the case of the ACs. We calculated probabilities for each coordinate using boundary conditions set by the eight RR~Lyrae stars located inside the SMCNOD. The bottom panel of Fig.~\ref{fig:SMCNOD-PRO-AC-RRL} depicts the distributions for $x$, $y$, and $z$ Cartesian coordinates with dashed lines representing a Gaussian distribution. The integration ranges (of RR~Lyrae stars in the SMCNOD) for each coordinate are denoted by red vertical lines. The resulting probability was then multiplied by the number of studied RR~Lyrae stars and divided by the number of assumed RR~Lyraes within the SMCNOD. Thus the probability of finding eight RR~Lyrae stars at the position of the SMCNOD is equal to 13.0\,\%. Therefore the probability is almost 20 times higher than the probability for finding four ACs. 

This finding seems to be consistent with the density profiles of the SMC (Fig.~\ref{fig:SMCNOD-DLSxyz-AC} and \ref{fig:SMCNOD-DLSxyz-RRLyr}). ACs at the possition of the SMCNOD slightly stand out and the probability of finding them at that position is very small, while RR~Lyrae stars smoothly follow the power-law for the SMC and we do not observe any deviation, which is in an agreement with the calculated probability, which is rather high for this region of the SMC. Therefore, RR~Lyraes seem to be evenly distributed across the SMC outskirts without forming any overdensity similar to the SMCNOD. Integration ranges and further details can be found in Table~\ref{tab:SMCNOD-Prob-table}. We note that the calculated probabilities comprise only stars that belong to the SMCNOD while the density profiles also contain stars that occupy the same bin range as the variables found in the SMCNOD. 

\setlength{\tabcolsep}{5pt}
\begin{table}
\centering
\caption{Table for properties used to calculate the probability of detecting variable stars at the position of the SMCNOD. Column 1 denotes the Cartesian components, columns 2 and 3 contain probability and boundary conditions for the given components based on the properties of the ACs in the SMCNOD. Columns 4 and 5 contain the same information as columns 2 and 3 but with respect to RR~Lyraes found in the SMCNOD. The last row represents the calculated probability for finding the same number of variables at the location and distance of the SMCNOD (4 ACs and 9 RR~Lyraes).}
\label{tab:SMCNOD-Prob-table}
\begin{tabular}{c|c|c|c|c|}
\multicolumn{1}{l|}{\multirow{2}{*}{}} & \multicolumn{2}{c|}{ACs}     & \multicolumn{2}{c|}{RR~Lyraes} \\  
\multicolumn{1}{|l|}{}                  & Prob. \%   & Boundary\,[kpc]        & Prob. \%   & Boundary\,[kpc] \\ \hline
\multicolumn{1}{|c|}{x}                & 8.68       & (2.78; 3.84)    & 67.80      & (-0.91; 3.68)   \\ \hline
\multicolumn{1}{|c|}{y}                & 0.58       & (7.01; 7.95)    & 0.13       & (6.27; 7.79)    \\ \hline
\multicolumn{1}{|c|}{z}                & 47.48      & (-1.15; 3.03)   & 38.89      & (-1.98; 2.22)   \\ \hline
\hline
\multicolumn{1}{|l|}{Total}            & \multicolumn{2}{c|}{0.69 \%} & \multicolumn{2}{c|}{13.0 \%}  \\ \hline
\end{tabular}
\end{table}

\section{Summary and conclusions} \label{sec:Conclus}

In this paper, we used variable stars from the OGLE-IV survey to trace and study the recently discovered stellar overdensity, SMCNOD, in the northern outskirts of the SMC. Using photometric data for the variables located at and near the SMC we studied their spatial, metallicity, and age distribution. We found that none of the CCs lie at the coordinates of the SMCNOD but prefer the more central star-forming regions of the SMC \citep[see also][]{Haschke2012bb}. The lack of young pulsators in the SMCNOD suggests that this overdensity does not contain stars younger than a few hundred million years consistent with the findings of \citet{Pieres2016SMCNOD} from a deep CMD of this region. In the R.A. vs. Dec plane, we found six EBs at the same coordinates as the SMCNOD. To verify their association with the SMCNOD, a thorough analysis of their infrared photometry and radial velocity curves or light curve modeling would be required.

We found four ACs coinciding with the coordinate and distance range of the SMCNOD. We calculated their ages, which range from 2 to 4.5\,Gyr. This agrees very well with the intermediate-age population scenario suggested by \cite{Pieres2016SMCNOD} for the SMCNOD. In addition, a similar age distribution is also seen in the SMC in general. Based on our performed tests, ACs occur rarely in the SMC outskirts and their high presence ($\sigma$ excess in density profile, see \ref{subsec:CCsDenProf}) at the location of the SMCNOD just by a chance is unlikely (less than 1.0\,\% probability, see \ref{subsec:probSMCNOD}). Therefore, we suggest that the ACs are indeed likely members of the SMCNOD overdensity.

We also found thirteen fundamental-mode RR~Lyrae stars at the coordinates of the SMCNOD. Using their period-metallicity-luminosity relation we were able to estimate their distances and separate those that lie in the foreground/background from those that most likely lie within the SMCNOD. From thirteen RRab-type stars sharing the same coordinates as the SMCNOD, eight lie at the same distance range as the SMC (within the errors), and thus probably lie inside the SMCNOD \citep[based on distance estimated by][]{Pieres2016SMCNOD}. The median metallicity of these pulsators is found to be around $\rm [Fe/H]_{\rm SMCNOD}=-1.71\pm0.21$\,dex as derived from the Fourier analysis of their light curves, which agrees well with the metallicity of the RRab variable population of the SMC $\rm [Fe/H]_{\rm SMC}=-1.77\pm0.48$\,dex \citep{Skowron2016}. The RR~Lyrae stars seem to be evenly distributed across the outskirts of the SMC and we do not see any excess in the density profile. In addition, the probability of finding eight of them at the location of the SMCNOD is higher than in the case of ACs (13.0\,\%, see Sec.~\ref{subsec:probSMCNOD}). We therefore conclude that these RR~Lyrae stars may, but need not be a part of the SMCNOD overdensity. They are consistent with being part of the extended old population of the SMC as such. That supports the CMD-based finding of \cite{Pieres2016SMCNOD} that an old population, if present in the SMCNOD, is small.

Overall, we showed that the SMCNOD can be traced with the AC stars that probably belong to an intermediate-age population and possibly also with old RR~Lyrae stars. The RR~Lyrae stars that may be members of the SMCNOD have a similar median metallicity as the SMC. Moreover, the similarity in the properties of the pulsating variables in the SMCNOD, SMC outskirts and in the SMC's main body suggests that the SMCNOD is not a remnant of a smaller, accreted satellite, given its inferred metallicity, which would be inconsistent with, for instance, the metallicity-luminosity relation for dwarf galaxies \citep[see, e.g.,][]{Grebel2003}. Furthermore, our calculated ages and metallicities rule out the potential merger of a primordial galaxy mentioned in \cite{Pieres2016SMCNOD}. The stellar population properties of the SMCNOD as indicated by its putative pulsating variable star members are consistent with those of the bulk SMC population. Hence the origin of the SMCNOD lies more likely in the SMC itself and may be possibly a somewhat older and less prominent counterpart of the luminous loop features in the outskirts of the SMC found and discussed by \cite{Vaucouleurs1972}, \cite{Bruck1980}, and \cite{Gardiner1992}, among others.


\section*{Acknowledgements}

Z.P. acknowledges the support of the Hector Fellow Academy. E.K.G and I.D. were supported by Sonderforschungsbereich SFB 881 "The Milky Way System" (subprojects A2, A3) of the German Research Foundation (DFG). RS is supported by the National Science Center, Poland, grant agreement DEC-2015/17/B/ST9/03421. We thank the anonymous referee for useful comments, which helped to improve the paper.









\appendix
\bsp	
\label{lastpage}
\end{document}